\documentstyle[12pt,epsfig,epsf]{article}

\def\be{\begin{equation}}
\def\bea{\begin{eqnarray}}
\def\ee{\end{equation}}
\def\eea{\end{eqnarray}}

\def\ra{\rangle}
\def\la{\langle}
\def\r{\right}
\def\l{\left}

\def\e{\eta}
\def\eps{\epsilon}
\def\m{\mu}

\def\a{\alpha}
\def\t{\tau}

\def\s{\sigma}
\def\b{\beta}

\begin{document}
\title{Relaxation, closing probabilities and transition from
  oscillatory to chaotic attractors in asymmetric neural networks} 
\author{Ugo Bastolla$^1$ and Giorgio Parisi $^2$}
\maketitle
\centerline{$^1$HLRZ, Forschungszentrum J\"ulich, D-52425 J\"ulich
  Germany}
\centerline{$^2$Dipartimento di Fisica, Universit\`a ``La Sapienza'', P.le Aldo
Moro 2, I-00185 Roma Italy}
\medskip
\centerline{Keywords: Disordered Systems, Attractor Neural Networks}

\begin{abstract}
Attractors in asymmetric neural networks with deterministic parallel
dynamics present a ``chaotic" regime at symmetry $\e\leq 0.5$ where
the average length of the cycles increases exponentially with system
size, and an oscillatory regime at high symmetry, where the average
length of the cycles is 2 \cite{Nut}.
We show, both with analytic arguments and numerically, that there is a
sharp transition, at a critical symmetry $\e_c=0.33$, between a phase
where the typical cycles have length 2 and basins of attraction of
vanishing weight and a phase where the typical cycles are
exponentially long with system size, and the weights of their
attraction basins are distributed as in a Random Map with reversal
symmetry. The time-scale after which cycles are reached grows
exponentially with system size $N$, and the exponent vanishes in the
symmetric limit, where $T\propto N^{2/3}$.
The transition can be related to the dynamics of the infinite system
(where cycles are never reached), using the closing probabilities as a
tool.

We also study the relaxation of the function $E(t)=-1/N\sum_i
|h_i(t)|$, where $h_i$ is the local field experienced by the neuron
$i$. In the symmetric system, it plays the role of a Ljapunov function
which drives the system towards its minima through steepest descent.
This interpretation survives, even if only on the average, also for
small asymmetry. This acts like an effective temperature:
the larger is the asymmetry, the faster
is the relaxation, and the higher is the asymptotic value reached.
$E$ reachs very deep minima in the fixed points of the dynamics, which
are reached with vanishing probability, and attains a larger value on
the typical attractors, which are cycles of length 2.

\end{abstract}

\section{Introduction}
The dynamics of Attractor Neural Networks with randomly distributed
synaptic couplings has been investigated in several works in recent
years, with particular attention to the effects of the asymmetry of
the synaptic couplings
\cite{P86,HGS,CS1,CS2,GRY,Hei1,Hei2,Hei3,Nut,NK,EO2,CFV,IM}. This
generalization is interesting not only because in the brain synapses are
not symmetric, but also because it introduces new complex features  in the
dynamics of such systems. At symmetry higher than a critical value the
system does never completely lose memory of its initial state
\cite{S-K,Hei3,EO2}, while at zero symmetry the
dynamics have the essential characteristics of chaos in continuous
systems even if the model has a finite state space \cite{CFV} (in this
case it has been shown analytically that chaos is
suppressed by strong enough thermic noise \cite{MSS}).

At low symmetry the average length of the attractors
increases exponentially with the size of the system, while for high
symmetry the typical attractors are cycles of length 2. This kind of
transition takes place also in other discrete disordered dynamical
systems \cite{K69}. The model that we consider here is very simple,
and a detailed study of this transition is possible. Moreover, the study of
asymmetry is interesting in the context of neural networks, both because 
it drops the unrealistic requirement of symmetric synaptic couplings formulated
in the classical Hopfield model \cite{Hop} and because the existence
of chaotic attractors can be a way to distinguish between a network
converging to a regular attractor since it has remembered a pattern
and a network which is in a confused state \cite{P86}. On the other
hand, the model that we study can also be seen as a modification of the mean
field model of Spin Glass proposed by Sherrington and Kirkpatrick
\cite{SK}, at temperature $T=0$ (if the couplings are not symmetric the
system is not Hamiltonian, but also in the case of symmetric couplings
the Hamiltonian of the system is different from the SK Hamiltonian due
to the parallel dynamics, see below).

The model that we consider consists in $N$ Ising neurons
$\s_i=\pm 1$ which are updated simultaneously according to
deterministic rules:

\be \s_i(t+1)={\rm sign}\l(\sum_j J_{ij}\s_j(t)\r). \label{map}\ee

The $J_{ij}$ (synaptic couplings) are quenched random
variables chosen from a distribution with average value zero and
variance $1/N$. The exact form of the distribution is irrelevant in
the infinite size limit, so we chose a two-valued distribution which
is easiest to implement numerically \cite{Hei3}. The degree of
symmetry of the synaptic couplings is parameterized by $\eta$, which
measures the correlation between $J_{ij}$ and $J_{ji}$:

\be \eta={\l\langle J_{ij}J_{ji}\r\rangle\over \l\langle
  J_{ij}^2\r\rangle}. \ee

If the limit $N\to\infty$ is taken before the limit $t\to\infty$, it
is possible to apply a Monte-Carlo method that reproduces exactly the
dynamics of a single spin in the infinite system \cite{EO}. This approach
\cite{EO2} and previous numerical works \cite{S-K,Hei3} show the presence of
a phase transition at $\e_{RM}=0.825$. At symmetry larger than
$\e_{RM}$ a system that was initially in a magnetized state
does never forget the initial configuration, and the
remanent magnetization, $m_\infty=\lim_{l\to\infty}
C(0,l)$, where $C$ is the correlation function, is different from
zero. At low symmetry the initial condition is completely
forgotten.

If the limit $t\to\infty$ is taken in a finite system, a different
situation is found. Since configuration space is finite and the
dynamics is deterministic, after a transient time the motion
takes place on periodic orbits. The length  of such orbits, the
size of their attraction basins and the number of orbits are random
variables, depending on the realization of the couplings $J_{ij}$. We are
interested in their scaling behavior as a function of system size, $N$. 
We can wonder what is the relation of
this situation with the dynamics of an infinite system, where
attractors do not exist. We will show that the properties of the
attractors can be deduced from the statistical properties of the
correlation function in the infinite system.
 
If the synaptic couplings are symmetric\footnote{
Another important model with symmetric couplings is the
Hopfield model \cite{Hop}, the prototype of Attractors Neural
Networks, where the $J_{ij}$ are given by the Hebbs
rule, $J_{ij}=\sum_{\m=1}^p\xi_i^\m\xi_j^\m$.}
({\it i.e.}, $J_{ij}=J_{ji}$), one can define \cite{Amit} the function 

\be E(t)=-\sum_{i,j}\s_i(t+1)J_{ij}\s_j(t)=-\sum_i\mid\sum_j
J_{ij}\s_j(t)\mid,\label{ener}\ee 
which is a non-increasing function of time. Since $E(t)$ attains a constant
value only on fixed points and on cycles of length 2, these are the
only attractors of the system.  On the fixed points the
function $E$ coincides with the Hamiltonian of the SK model of Spin
Glass \cite{SK}, $H=-\sum_{ij}J_{ij}\s_i\s_j$, thus the fixed points
of the symmetric networks are also metastable states for the SK
model. The typical attractors are nevertheless cycles of length 2:
nearly every initial point (in the limit $N\to\infty$)
converges to a cycle of length 2. For symmetry reasons, the value of
$H$ is in average zero in such cycles, but $E$ has a low value. We
observed that the lowest values of $E$ are attained on the fixed
points, which are reached with vanishing probability. Thus, under the
point of view of relaxation, cycles of length 2 act as traps.

Let us summarize our results about non-symmetric synaptic
couplings. In this case, $E(t)$ may increase and it is not anymore a
Ljapunov function. Nevertheless, we saw that its average value is a
non-increasing function also for asymmetric couplings.
Cycles of every length may exist, but, if the asymmetry
is small, the typical cycles are still cycles of length 2. Each of
them has an attraction basin of vanishing weight (so that the average
weight of the attraction basins is zero), but the sum of such weights
tends to 1 in the infinite size limit. This situation persists up to
$\e_c\approx 0.33$, where the sum of the weights of cycles of length 2
suddenly drops to 0 in the infinite size limit. The
typical attractors are now very long cycles,
whose length increases exponentially with system size (chaotic phase).
The number of such cycles is much smaller than the number of cycles of
length 1, but the average weight of their attraction basins is finite,
like in the Random Map model \cite{DF}. Indeed, we argue that the distribution
of the weights is the same as in a Random Map model with reversal
symmetry, as in the limit case $\e=0$ \cite{BP1}\footnote{
The case $\e=0$ is in some sense peculiar, since the average number of
attractors increases linearly with system size \cite{eta0}, as in a
Random Map and in contrast with the case $\e\neq 0$, where the number
of fixed points increases exponentially with N \cite{GRY}.}.

Between $\e=0.5$ and $\e=0.33$ long cycles are still a negligible portion of
phase space, but the average length is dominated by the tails of the
distribution, and increases exponentially with system size, in
agreement with the results of N\"utzel \cite{Nut}. However, we prefer
to place the transition between the two 
regimes at  $\e_c=0.33$, where the nature of the typical attractors
abruptly changes. We also measured the typical transient time (the time
necessary to reach a cycle). It appears from our data, in contrast
with previous numerical results, that the transient time
increases exponentially with system size for $\e<1$, and as $N^{2/3}$
for $\e=1$. For $\e\approx 1$ there are two regimes: a power law at
small $N$ (roughly with the same exponent 2/3 as in the case $\e=1$)
and an exponential increase for larger systems, with an exponent
vanishing in the limit $\e\to 1$. This behavior is in agreement with
our theoretical expectations (see below).

The above observations hold for $\e>0$. The case $\e<0$, which
corresponds to antisymmetric couplings, is related in a simple way to
the symmetric case in the infinite size limit, as we shall see. In
this case the phase space is dominated by cycles of length 4 for
$\e\leq -0.33$, while exponentially long cycles prevail for larger
values of $\e$. The transition point is $\e_c=-0.33$.

We study, numerically and
in part analytically, the closing probabilities, that express the
probability that a trajectory not yet closed closes on a cycle of
length $l$ after a time $t+l$. The closing probabilities are nothing but
the tail of the distribution of the correlation function
$q(t,t+l)$, thus they establish a link between the properties of the
infinite system and the properties of the attractors \cite{BP0}.
In this framework, the transition can be understood in this way: since
the distribution of $q(t,t+2)$ is peaked around a much larger value
than for any other value of $l$, the closing probability on cycles of
length 2 is much larger than for any other cycle, for a factor
exponentially large with system size. This effect decreases as $\e$
decreases, and, at a critical parameter, it is not enough to balance
the large number of possible long cycles, which then prevail.

The paper is organized as follows: in section 2 we study analytically
the properties of the attractors under some hypothesis on the closing
probabilities. We derive a
condition on the exponent of the closing probability that
corresponds to the dynamical transition.  We argue
that the distribution of the weights of the attraction basins is
the same as for a Random Map with reversal
symmetry in the whole chaotic phase, while the typical weights tend to
zero in the oscillatory phase. Thus also this quantity has a
discontinuity at the transition. In section 3 we present
our numerical results, reporting the properties of the
attractors, the relaxation of $\overline{E(t)}$, the distribution of
the correlation functions and the closing probabilities. In
the final discussion we point out possible extensions of this study.

\section{Closing probabilities and attractors}

The natural distance in the configuration space of the system is the
Hamming distance, that measures the number of elements in a
different state in two configurations.
An equivalent information is given by the correlation
between configurations, or overlap. Let us consider the overlap
between two configurations at different times along the same trajectory:

\be q(t,t+l)={1\over N}\sum_i \s_i(t)\s_i(t+l).\ee

This is a random variable, depending a) on the quenched disorder
(dynamical rules), and b) on the randomly chosen initial point.
Knowing its distribution, we can reconstruct all the properties of the
attractors. In particular, we have to compute
the probability that $q(t,t+l)$ is equal either to 1 or to -1. The
first case corresponds to a trajectory that,
after a transient time $t$, enters a limit cycle of length $l$.
The case $q(t,t+l)=-1$ corresponds to a trajectory that, after a transient.
time $t$, enters a cycle of length $2l$. In fact, if $C(t)$ has been
reversed after a time $l$, also $C(t+l)$ shall be reversed after a
time $l$, since the map (\ref{map}) commutes with the reversal
operator $\cal R$ defined by ${\cal R}\{\s_1, \cdots \s_N\}=\{-\s_1,
\cdots -\s_N\}$, thus $C(t+2l)$ will be equal to $C(t)$.

We are interested in the first time when a trajectory ``closes"
visiting a configuration already attained. Thus, we have
to impose the condition that no configuration has yet
been repeated before the time $t+l$. The closing probabilities are then
conditional probabilities:

\bea & & \pi_N^+(t,t')=\Pr\l\{q(t,t')=1\mid A_{t'}\r\}, \\
& & \pi_N^-(t,t')=\Pr\l\{q(t,t')=-1\mid A_{t'}\r\}, \nonumber\eea
where the symbol $A(t')$ represents the condition that it never happened,
before time $t'$, either $q=1$ or $q=-1$.

\subsection{The overlap in an infinite system}
Taking this opening condition into account reconciles
the apparent discrepancy between the behavior of a finite system
\cite{NK} and of an infinite one \cite{EO2}. In an infinite system, the
average\footnote{ 
Here, as it is usual in the theory of disordered systems,
the over-line denotes the average respect to the disorder, while the
angular brackets denote an average over the initial configurations
that in this case is not needed \cite{GDM}.}
overlap after two time-steps,
$Q(t,t+2)=\overline{ q(t,t+2)}$, tends, as $t\to\infty$, to an
asymptotic value that is an analytic function of $\e$. In contrast, in
a finite system, if time is long enough a limit cycle is reached. Then
$Q(t,t+2)$ tends to 1 with probability 1 in the oscillatory phase
while it is less than 1 in the chaotic phase.

Because of the opening condition we can use, as a first
approximation, the properties of the correlation function in an
infinite system to derive the properties of the attractors in a finite
system. We list here some properties of the conditional distribution
of $q(t,t+l)$ which we expect to hold if the limit $N\to\infty$ is taken with
$t$ and $l$ fixed. In what follows, the opening condition
is always assumed to hold. 

We consider first $\e\geq 0$. Qualitatively, we
expect that the behavior of the overlap at $-\e$ is related in a
simple way to the one at $\e$. A simple argument runs like this: all the
couplings where $J_{ij}J_{ji}$ is positive tend to align $\s_i(t)$ 
and $\s_i(t+l)$, so that for $\e>0$ there is an effective ferromagnetic
interaction between the time slice at time $t$ and the time slice at
time $t+l$, and an effective antiferromagnetic interaction for
$\e<0$. These interactions have a different sign, but the same
strength. They determine the main features of the overlap
distribution: for instance, the variance and the non-vanishing value
of $Q(t,t+l)$ (for even $l$; for odd $l$ the distribution is symmetric)

\begin{enumerate}
\item The overlap distribution depends exponentially on system size:

\be \Pr\{q(t,t+l)=q\}\approx {A_{t,t+l}(q;\e)\over\sqrt N}
\exp\l(-N\a_{t,t+l}(q;\e)\r). \label{exp}\ee

This is a kind of finite entropy statement, and it can be understood
as follows: the overlap is the average of $N$ terms. They are not
independent, but their correlations are small enough so that the variance
of the overlap is of order $1/N$. Thus, in analogy with the Shannon-Mc
Millan theorem, we expect (\ref{exp})\footnote{
Another way to see this equation is to compute the probability of
$q(t,t+l)$ with the dynamical functional integral.}.

\item Weak time translation invariance: in the limit $t\to\infty$, the
overlap $q(t,t+l)$ converges to a well-defined limit distribution.
We call $\overline{\a_l}(\e)$ the limit value of $\a_{t,t+l}(q=1,\e)$.

\item Symmetry: for $l$ odd, the distribution of $q(t,t+l)$ is symmetric
around $q=0$. This can be easily proved using the inversion symmetry
$J\to -J$ of the distribution of the synaptic couplings. It implies
that $Q(t,t+l)=0$ for $l$ odd, and, more in general,
$\a_{t,t+l}(q;\e)=\a_{t,t+l}(-q;\e)$.
 
The variance of the distribution is larger than $1/N$ (the value for
a binomial distribution), since there is a positive correlation
between $\s_i(t)\s_i(t')$ and $\s_j(t)\s_j(t')$ if $J_{ij}$ and
$J_{ji}$ have the same 
sign. The closing probability is accordingly larger than
$2/2^N$. Both the variance and the closing probability decrease with
$l$ and increase with $\e$.

\item For $l$ even, the most probable overlap, $Q(t,t+l)$, is positive
  and decreases with $l$ (correlations
decay in time). In the phase where there is not remanent magnetization
($\e<0.825$) the limit value is zero, and $q(t,t+l)$ reaches
asymptotically a symmetric distribution (the limit of large
$t$ having been taken in advance). The closing
probability consequently decreases with $l$ and increases with $\e$.

\item  The limit distribution reached in the limit  $l\to\infty$ is
  found numerically to be the same both for even and odd
  $l$. We denote the exponent of the closing probability by the symbol
  $\a_{t,\infty}(\e)$ (since the limit distribution is symmetric, the
  exponent is the same both for closing with $q=1$ and for closing
  with $q=-1$).

\end{enumerate}

As a consequence, the closing probability has a maximum for $l=2$ at
fixed $t$, and it is exponentially larger than for any other value of
$l$. Thus cycles of length 2 are found with the highest frequency.

\vspace{.5cm}
In antisymmetric networks ($\e<0$) the situation is slightly more
complicate: we have to distinguish $l$ odd, $l=4m$ and $l=2(2m+1)$.

\begin{itemize}
\item For odd $l$ the average overlap is zero and the variance is
  smaller than in a binomial distribution, since the correlation
  between $\s_i(t)\s_i(t')$ and $\s_j(t)\s_j(t')$ is negative if $J_{ij}$ and
  $J_{ji}$ have opposite sign. Thus the closing probability is
  smaller than in the case of a binomial distribution and increases
  as $l$ increases. Accordingly, cycles of odd length
  are very rare in antisymmetric networks.

\item If $l=2(2m+1)$ ($l=2,6,10,\cdots$) there is an effective
  antiferromagnetic interaction between times slices $t$ and
  $t+l$. Thus $Q(t,t+l)$ is negative and the trajectories close
  preferentially with $q(t,t+l)=-1$ producing cycles of length
  $2l$.

\item If $l=4m$ the effective interaction is ferromagnetic, and the
  behavior is the same as for positive $\e$. 
\end{itemize}

In the case of negative $\e$, we expect $\pi_N^-(t,t+l;-\e)\approx
\pi_N^-(t,t+l;\e)$ for even $\e$, asymptotically in $N$. Thus the
dominant closing probability is $\pi_N^-(t,t+2)$, and the most
frequent cycles are cycles of length 4. Since the transition regarding
the attractors is governed by the comparison between the closing
probability with $l=2$ and that with $l=\infty$ (see below),
the transition for negative $\e$ should take place at $\e'_c=-\e_c$.
Simulations confirm very well these arguments, so that in
most of what follows we limit our study to the case $\e>0$. 

\vspace{.5cm}
In finite systems the situation is more complicate, and
some of these properties do not hold when the times $t$
and $l$ are large respect to system size. For instance, in this case
$q(t,t+l)$ does not reach an asymptotic distribution in the limit
$t\to\infty$ : it is easy to see that its average value has to
decrease at large $t$ as a consequence of the opening condition.
In fact, $\tilde\pi_N(t')=\sum_t \pi_N(t,t')$ (integrated closing
probability) can not exceed 1, since it represents the probability
that a trajectory closes at time $t'$. Thus $\pi_N(t,t+l)$ must
finally decrease with $t$ and the conditional distribution of the
overlap $q(t,t+l)$ can not become really stationary in $t$.
However, as $N$ increases this effect becomes weaker and weaker. 

We are interested in the closure of the cycles, which takes place at a
time-scale $\t$ exponentially increasing with $N$, and it is not {\it a
  priori} clear whether we can neglect these finite size effects 
or not. However, the predictions that can be drawn from this
simplified description capture the essential features observed in the
simulations. We derive in the following these predictions.

\subsection{Cycles of length 2}
We compute now the  probability that a trajectory closes on a cycle of
length $l=2$. Let us consider $\e>0$. We
want to show that, if $\bar{\a}_2(1;\e)<{1\over
  2}\bar{\a}_\infty(1;\e)$, then all trajectories reach
attractors of length 2 with probability 1 as $N\to \infty$. We call
this situation the oscillatory phase.

It is very easy to go from the closing probabilities to the
probability distribution of cycle lengths. The probability that the
trajectory has not closed up to time $t$ is given by

\be P_N(t)\approx
\exp\l(-\sum_{t'=1}^t\sum_{l=1}^{t'}\pi_N(t'-l,t')\r), \ee
so the time-scale with which this quantities decays gives the
time-scale of typical closing events.

We introduce the function $f_N(t)$, which expresses
the ratio between the probability that a cycle of length
$l\neq 2$ is reached at time $t$ and the probability that a cycle of
length 2 is reached at the same time:

\be f_N(t;\e)={\sum_{l\neq 2}^t \pi_N(t-l,t)\over
  \pi_N(t-2,t)}. \label{f}\ee

In the above hypothesis, this is an increasing function of $t$ when $t$
is large and the overlap distributions have reached the
asymptotic value.
The probability that a cycle of length 2 is reached at time $t$,
under the condition that a cycle was reached, is given by
$\l(1+f_N(t)\r)^{-1}$. This is very high for $t$ small and decreases
with time, since more possible cycles enter the game. 
Let us compute this quantity at the time scale $\t_2=e^{N\a_2}$:
\be f_N\left(\t_2;\e\right)\approx
\exp\l(N(2\a_2-\a_\infty)\r), \ee
where
$\a_2$ is the exponent of the asymptotic closing probability for
cycles of length 2. Two
different situations occur:

\begin{itemize}
\item 
$\a_2<\a_\infty/2$: oscillatory phase. $f_N(t)$ is
exponentially small up to times of order $\t_2=e^{N\a_2}$. But at such
times the probability that a trajectory is not yet closed, $P_N(t)$,
is very small: $P_N(t)\leq \exp(-t/\t_2)$. Thus, for very large
sizes $N$, almost all the cycles close before this time, and they are
cycles of length 2.

\item 
$\a_2>\a_\infty/2$: chaotic phase. In this case, the time scale
at which most of the trajectories are close is proportional to
$\t_\infty=e^{\a_\infty/2}$, since it holds $P_N(t)\leq
\exp\l(-(t/\t_\infty)^2\r)$. At this time, $f_N(\t_\infty)$ is already
very small and the probability that at least a cycle of length 2 is
reached is vanishingly small with $N$.
\end{itemize}

Since $\a_2(\e)$ tends to zero for $\e\to 1$ (as we noted above, in
this case the most probable value of the overlap is $Q(t,t+2)=1$ and
the distribution is not exponential), while $\a_\infty$ does not
vanish in this limit (actually, for $\e=1$ the value of $\a_\infty$ is
infinite, since only cycles of length 1 and 2 are found), and for
$\e=0$ the exponents $\a_2$ and $\a_\infty$ are roughly equal, it must
exist a critical value of $\e$ at which the condition
$\a_\infty=2\a_2$ is fulfilled.

\subsection{Transient times}
According to the above argument, the typical closing times are given by

\be {\rm Tr}\propto \exp\l(N\min(\a_2,\a_\infty/2)\r), \label{Trexp}\ee
and the argument predicts that transient time increases exponentially with
system size, with an exponent that vanishes in the limit $\e\to
1$. Close to this value, very large systems are needed in order to
distinguish this behavior from a stretched exponential or a power
law. For $\e=1$, the most likely value of $q(t,t+2)$ is $q=1$ and the
exponent $\a_2$ vanishes. Thus the closing probability has to be
computed before $q(t,t+2)$ reaches its stationary distribution. In the
Gaussian approximation,

\be \pi_N(t,t+2;\e=1)
\propto{1\over\sqrt N}\exp\l( -{(1-Q_2(t;\e=1))^2\over 2
NV_2(t)}\r) \approx {1\over\sqrt N}\exp\l(-{(1-Q_2(t;\e=1))\over 2
N}\r), \ee
since, for $Q_2$ close to 1, the variance goes to zero
as $V_2\propto 1-Q_2$. It is known from previous numerical studies
\cite{GDM} that $1-Q_2(t,\e=1)\propto t^{3/2}$. Thus the
time-scale at which trajectories close for $\e=1$ is a power-law of
system size: 

\be {\rm Tr}(\e=1)\propto N^{2/3}. \label{Trpow}\ee

This prediction is in good agreement with the numerical results in
N\"utzel \cite{Nut} and in the present work. When $\e$ is close to 1,
there are two regimes: at small $N$, the trajectories close when the
overlap distribution is not yet stationary. Since $Q(t,t+2)$ tends to
its stationary value $Q_2^*(\e)$ as $t^{-2/3}$ independent of $\e$
\cite{Hei3,EO2,Nut}, we expect and observe also in this case that ${\rm
  Tr}\propto N^{2/3}$. At larger size the overlap becomes stationary
by the time when cycles close, and the exponential dependence shows
up, with an exponent that, according to the Gaussian approximation, is
given by

\be \a_2(\e)\propto {1-Q_2^*(\e)\over 2}. \ee

Numerically, it is found $\a_2(\e)\propto (1-\e)^2$.

\subsection{Average length of the cycles}
At the time $t\propto\exp(N\a_2)$ when typical cycles of length 2 close the
probability that a long cycle close goes as
$\exp\l((\a_2-\a_\infty)N\r)$ and the average length of cycles longer
than 2 grows as most as $\exp\l((3\a_2-\a_\infty)N\r)$. Thus, for
$\a_2<\a_\infty/3$, the typical length of the cycles and the average
length coincide, and they are equal to 2.
At lower symmetry very large cycles appear with a probability
that, though vanishing, is large enough to let them dominate the
average length. At this point the average length of the cycles
increases exponentially with system size. This change takes
place inside the oscillatory phase, in which the typical cycles have
length 2. We prefer to use the probability $P(2)$ to find a cycle of
length 2 as the order parameter to describe the transition, instead of
using the logarithm of the average
length of the cycles, which is dominated by the tails of the distribution. 

Our simulations show that the transition defined by $P(2)$ happens at
$\e_c\approx 0.33$, a value much smaller than the threshold where
the remanent magnetization vanishes, $\e_{RM}=0.825$, and also smaller
than the threshold at which the average length of the cycles starts to
increase exponentially, $\e_L=0.50$ \cite{Nut}.

\subsection{Weights of the attraction basins}
Another quantity that can be used as an order parameter is the average
weight of the attraction basins.
The weight $W_\a$ of the attraction basin of cycle $\a$ represents the
probability that a random chosen configuration reaches ultimately this
cycle. The $\{W_\a\}$'s are random variables, depending on the
realization of the couplings. It is convenient to define the moments
of their distribution \cite{DF1}:

\be \overline{Y_n}=\sum_\a \overline{W_\a^n}. \label{Yn}\ee

$Y_1$ is equal to 1 because of the normalization. The average weight
of the attraction basins, $\overline{Y_2}$, is equal to the
probability that two randomly chosen trajectories reach the same
attractor \cite{DF1}. Let us evaluate such a probability. We define
$\pi_N^{(2)}(t,t')$ as the probability that the configuration at time $t$ on
the first trajectory is equal to the configuration at time $t'$ on the
second trajectory, given that the initial configurations are chosen at
random. This is, asymptotically in $t$, equal to $\exp(-N\a_\infty)$,
if the remanent magnetization is zero, or smaller, otherwise. In fact,
we expect that the overlap of two independent trajectories is asymptotically
equal to the overlap between configurations at long time distance
along the same trajectories, if the dynamics looses memory of the
initial state (this requires that the remanent magnetization
vanishes), and is smaller in the case in which the memory of the
initial state is not lost. Predictions based on this expectation are
well satisfied in other disordered dynamical systems like Kauffman
networks \cite{BP0} and discretized chaotic maps \cite{BP2}. 

Thus we expect that the time when two trajectories in the same
attraction basin eventually meet grows
as $\exp(-N\a_\infty/2)$. In the oscillatory phase this is much
larger than the time scale over which a trajectory reaches a cycle,
$\exp(-N\a_2)$, so that we expect that the probability that two
trajectories meet before reaching their limit cycles tends to
zero. This is confirmed by the simulations.

In the chaotic phase the two time-scales are equal, and we expect
non-vanishing $\overline{Y_n}$. Moreover, we expect that their values
coincide with the moments of the weights of the attraction basins of a
Random Map model \cite{DF}, endowed with reversal symmetry, like in
the previously studied case $\e=0$ \cite{BP1}. The computation follows
exactly the same steps as in that
case: in fact, the Random Map distribution follows only from
the facts that the closing probabilities $\pi_N(t,t+l)$ tend to an
asymptotic limit independent on both $t$ and $l$, and that the
``transient" regime (small $l$) does not contribute significantly to
the closure of the cycles. The result of the computation is \cite{BP1}:

\be \overline{Y_{n+1}}={1\over 2}{(n!)^2\over (2n+1)!}\l(4^n+2^n\r).\ee

For instance, $\overline{Y_2}=1/2$, $\overline{Y_3}=1/3$,
$\overline{Y_4}=9/35$... Our numerical results are
consistent with this prediction (see Fig. \ref{fig_phase}), although,
for $\e>0.1$, $\overline{Y_2}$ is still far from the asymptotic regime
for all the systems that we could simulate. 

\subsection{Average number of cycles}
We conclude the discussion of the properties of the attractors showing
that the closing probabilities allow to compute the average number of cycles
of length $l$ through the relation

\be \overline{N_c}(l)= 2^N \l({1\over
  l}\pi_N^+(0,l)+{2\over l}\pi_N^-(0,l/2)\r) \ee
(the last term is present only for even $l$).
Thus the number of cycles varies exponentially with $N$.
For odd $l$ and $\e>0$, $\a_{0l}(q=1)$ is smaller than $\ln 2$ (in fact,
there is a positive correlation between $\s_i(0)\s_i(l)$ and
$\s_j(0)\s_j(l)$ if $J_{ij}$ and $J_{ji}$ have the same sign. The
correlation change of sign in case of negative $\e$, and the number of
odd cycles becomes exponentially small). For even $l=2m$,
$\a_{0l}(1)$ is also smaller than $\ln 2$ (in fact, the average value of
$\s_i(0)\s_i(l)$ is different from zero). The two cases are related:
for instance, the number of cycles of length 2 increase as
the square of the number of cycles of length 1 \cite{GRY}. With our
notation, this relation reads $\a_{01}-\a_{02}=\ln 2$.


For large $l$, no matter if even or odd, if the remanent magnetization is
equal to zero, the distribution of $q(0,l)$ tends to a
binomial distribution with average value 1/2. Thus the
average number of cycles of length $l$ tends to 1 for large $l$.
A funny consequence of a non-vanishing remanent
magnetization is that in this case the number of cycles of
length $l$ increases exponentially with a non-vanishing exponent for
every value of $l$.

\section{Numerical results}
\subsection{Properties of the attractors}
In order to monitor the transition from short cycles to long
attractors, we measured 5 quantities: the probability that
a cycle of length 2 is reached, $P(2)$, the average weight of the
attraction basins $\overline{Y_2}$ (equation \ref{Yn}), the average
length of the cycles and of the transient time, and the distribution of
cycle length.

\begin{figure}
  \begin{minipage}{77truemm}
    \begin{center}    
      \epsfig{file=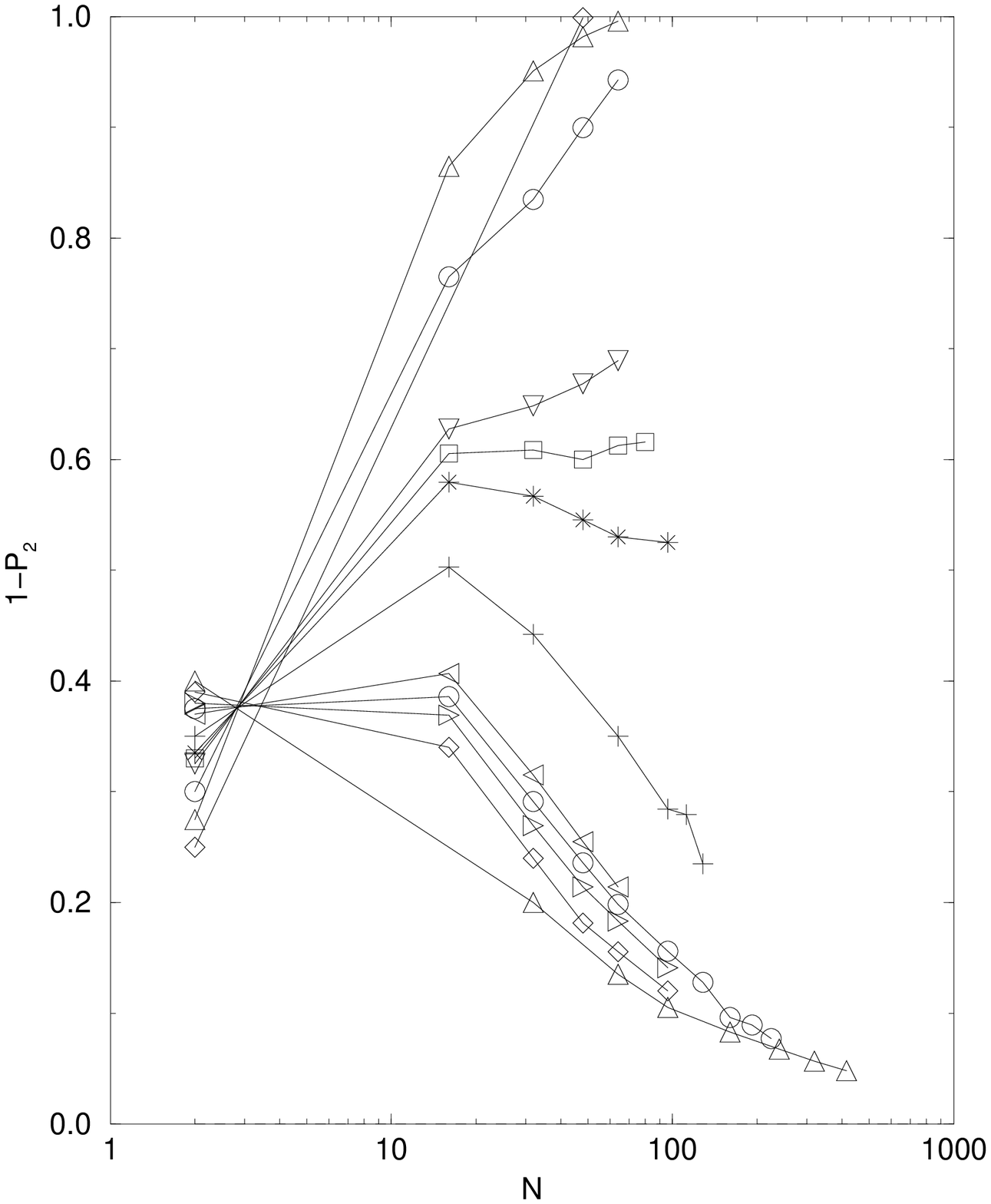,width=75truemm}
    \end{center}
  \end{minipage}
  \begin{minipage}{77truemm}
    \begin{center}    
      \epsfig{file=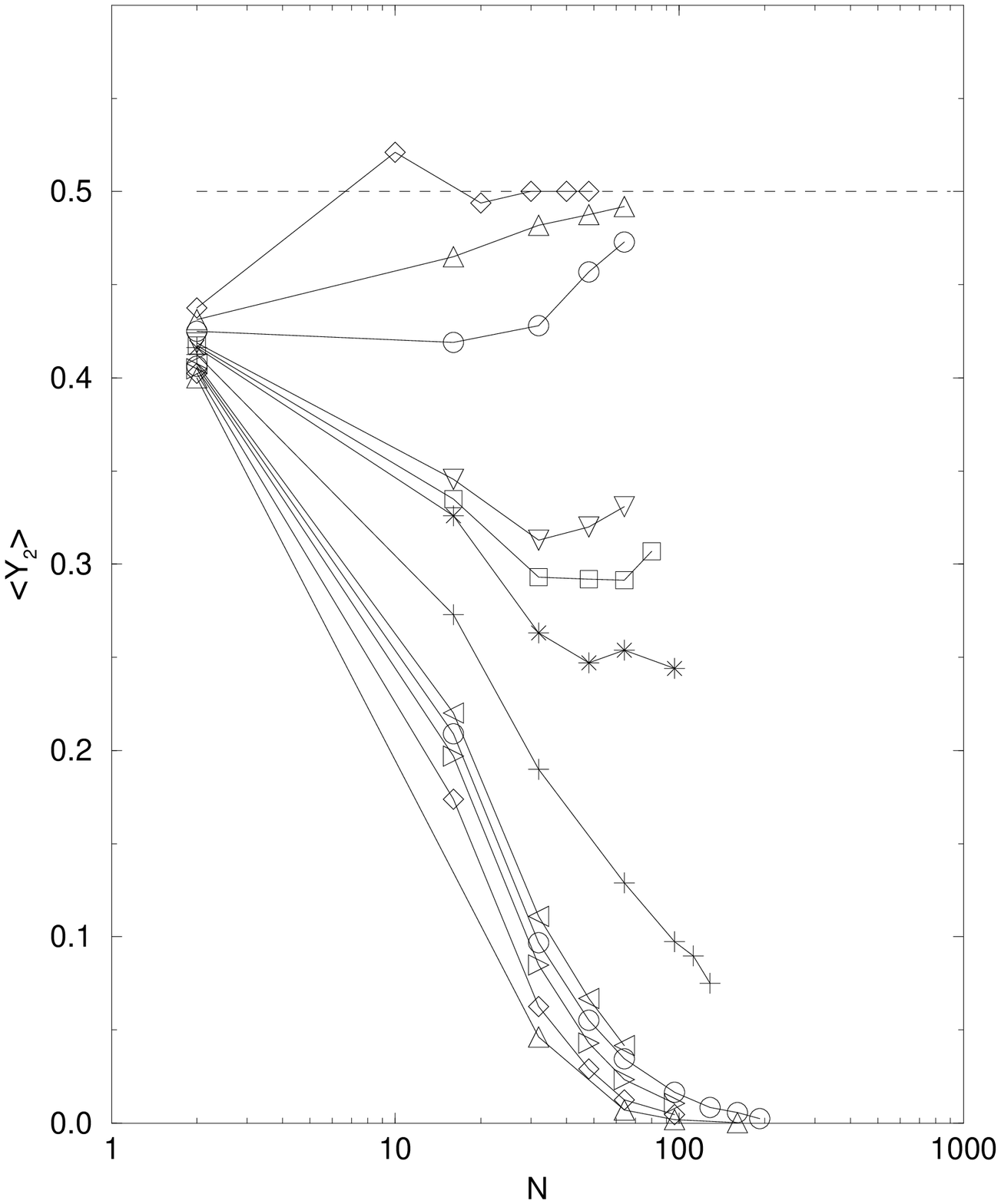,width=75truemm}
    \end{center}
  \end{minipage}

  \caption{ (a): Probability to find a cycle of length different from 2,
    and (b): Average weight of the attraction basins, as a function
    system size. From top to bottom, the values of the symmetry is
    $\e=0$, $0.1$, $0.2$, $0.3$, $0.32$ (squares), $0.34$ (stars),
    $0.4$, $0.48$, $0.5$, $0.52$, $0.56$, $0.6$. The error bars are,
    in the worst cases, comparable with system size.}
  \label{fig_phase}
\end{figure}

\vspace{.3cm}
The probability to reach a cycle of length 2, $P_2$, initially
decreases with system size. For $\e\geq 0.34$, it reaches a minimum
and starts to increase, going asymptotically to the value 1. For
$\e\leq 0.32$ it seems that it always decreases with system size,
apparently going to zero
(though we can not exclude that it starts to increase at a size larger
than the ones that we could simulate). Thus we place the transition at
$\e_c\approx 0.33$ (see Fig. \ref{fig_phase}a), remarking that the
threshold could be overestimated.

The average weight of the attraction basins has also a non-monotonic
behavior with system size, but it leads to a different systematic
error: it starts decreasing with system size, then reaches a
minimum and eventually increases for $\e\leq 0.32$ , probably tending
to the value $0.5$ typical of a Random Map with reversal symmetry
(dashed line in Fig. \ref{fig_phase}b). For $\e\geq 0.34$
$\overline{Y_2}$ apparently tends to vanish, although we can not
exclude that it starts to increase at a larger size. Thus also from
this measure we estimate $\e_c\approx 0.33$, but this time the
threshold could be underestimated. From this measurement and the previous
one, we conclude then that $\e_c=0.33\pm 0.01$. 

We fitted the average length of the cycles and the average transient
time with an exponential function of system size, $L\simeq
\exp(\a_L(\e)N)$ and $T\simeq\exp(\a_T(\e)N)$ respectively. In the
first case, the fit is good for $\e\leq 0.5$, and the exponent
$\a_L(\e)$ vanishes at this point. For larger symmetry, the average
length of the cycles tends to 2, but not in a monotonic way: it starts
increasing, reaches a maximum and then decreases. Thus we can not
exclude that the exponential increase of the cycles stops at a
smaller value of the symmetry parameter if larger systems are
considered.

The average transient time increases as $\exp\l(N\a_T(\e)\r)$ for
$|\e|<1$, and $\a_T(\e)$ vanishes at $\e=1$ as $(1-\e)^{2.0\pm 0.1}$,
as expected. At $\e=1$ the transients grow as a power law of system
size, $Tr\propto N^b$, with $b=0.66\pm 0.01$, in agreement with the
prediction of equation \ref{Trpow} and the previous numerical work by
N\"utzel \cite{Nut}. At $\e=0.9$ we found two
regimes: for $N\leq 512$, the transient time increases as a power law
with exponent $b=0.66$ as in the system with $\e=1$. At $\e>1000$ the
transient time increases faster with $N$. The analysis of the closing
probability, which is a less noisy measure, allow one to conclude that
the behavior with $N$ is, asymptotically, exponential with a very
small exponent. The first regime reflects the time needed to reach the
asymptotic distribution of the overlap, as discussed in section 2 for
the case $\e=1$, as predicted in the previous section.

\begin{figure}
  \centerline{
    \epsfysize=8.0cm 
    \epsfxsize=10.0cm 
    \epsffile{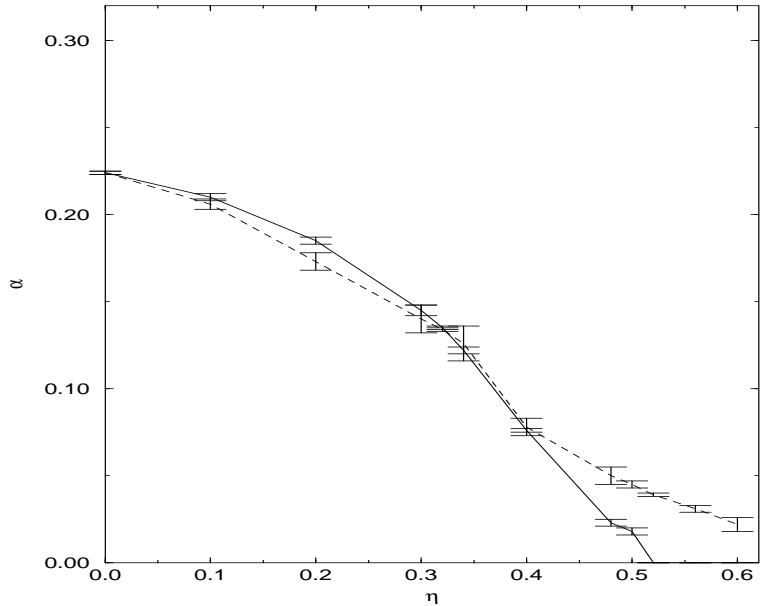}}
\caption{ Exponent of the average cycle length (full line) and of the
  average transient time (dashed line) as a function of the asymmetry.}
\label{fig_alp}
\end{figure}

We plot in Fig. \ref{fig_alp} the exponents $\a_L(\e)$ and
$\a_T(\e)$, for $\e\geq 0$. We checked for $\e=-0.4$ that,
asymptotically in system size, it holds $\a_L(-\e)\approx\a_L(\e)$ and
$\a_T(-\e)\approx\a_T(\e)$.

\vspace{.5cm}

The distributions of the lengths of the cycles have different
features in three different regions. As a general rule, we have to
consider separately cycles of even length and cycles of odd length,
not only because the first ones can be obtained in two different ways
(they close with $q(t,t+l)=1$ and $q(t,t+l/2)=-1$) but also because
the closing probability that $q(t,t+l)=1$ is larger for even
$l$. That's what we observed:

\begin{figure}
  \centerline{ 
    \epsfysize=8.0cm 
    \epsfxsize=10.0cm 
    \epsffile{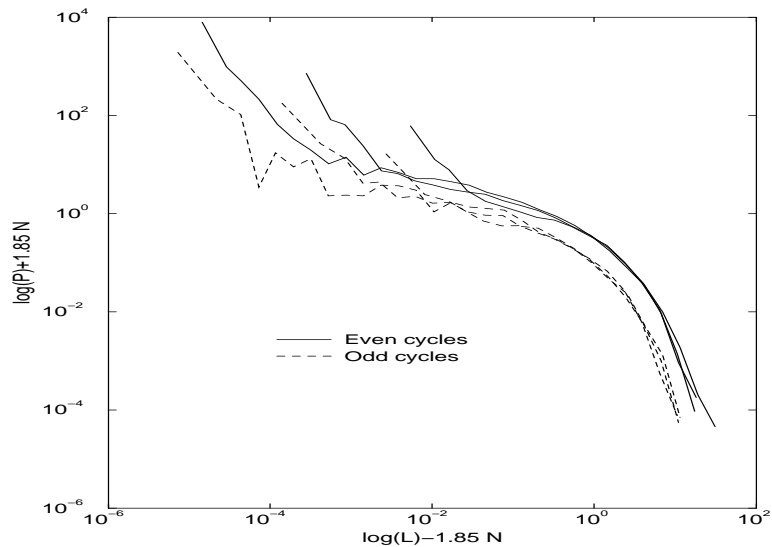}}
    \caption{Probability distribution of the rescaled length of the cycles
     in the chaotic phase ($\e=0.2$). $N=32$, 48 and 64. Systems of
     different sizes have been rescaled.}
\label{fig_per60}
\end{figure}

\begin{figure}
  \begin{minipage}{77truemm}
    \begin{center}    
      \epsfig{file=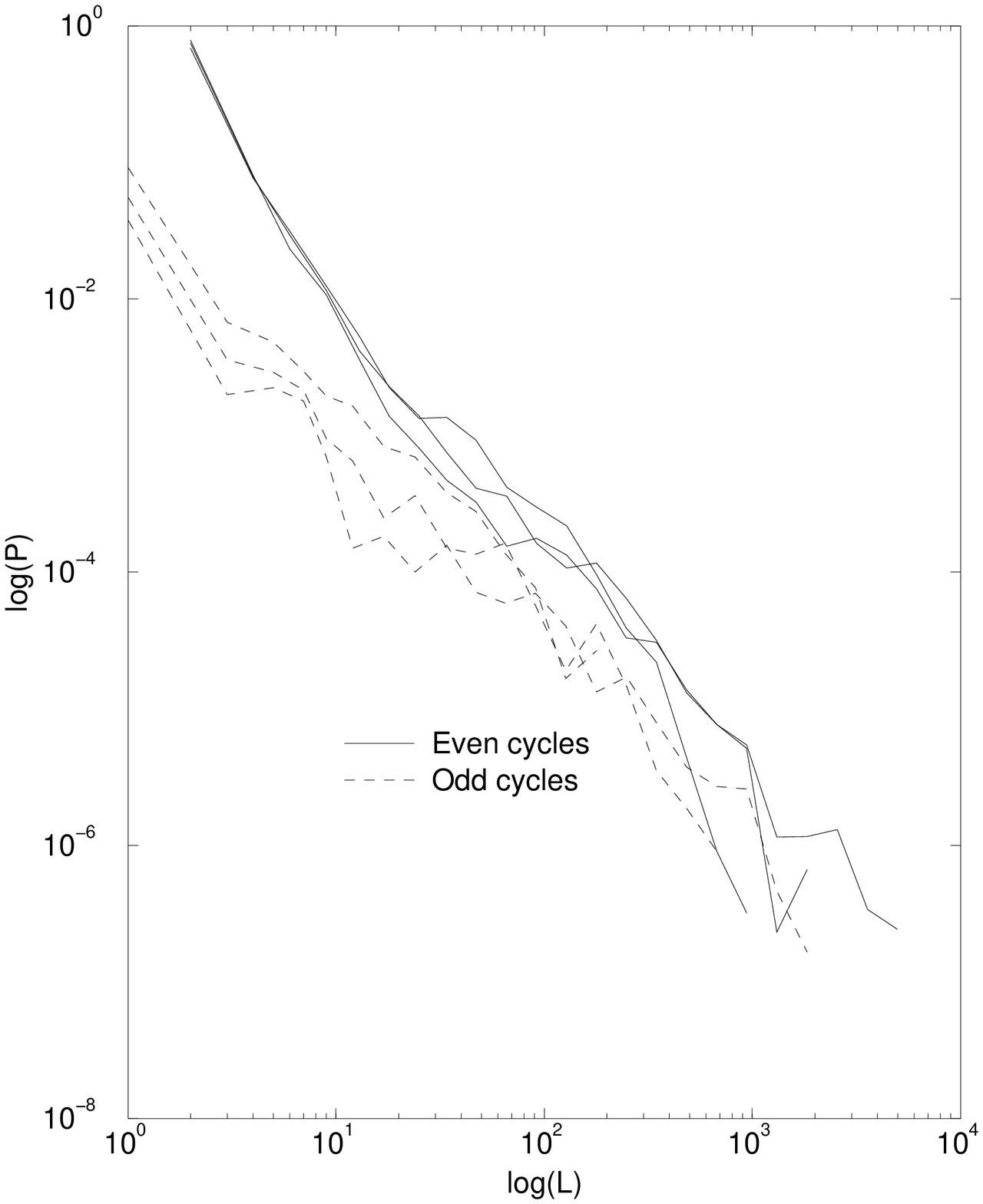,width=75truemm}
    \end{center}
  \end{minipage}
  \begin{minipage}{77truemm}
    \begin{center}    
      \epsfig{file=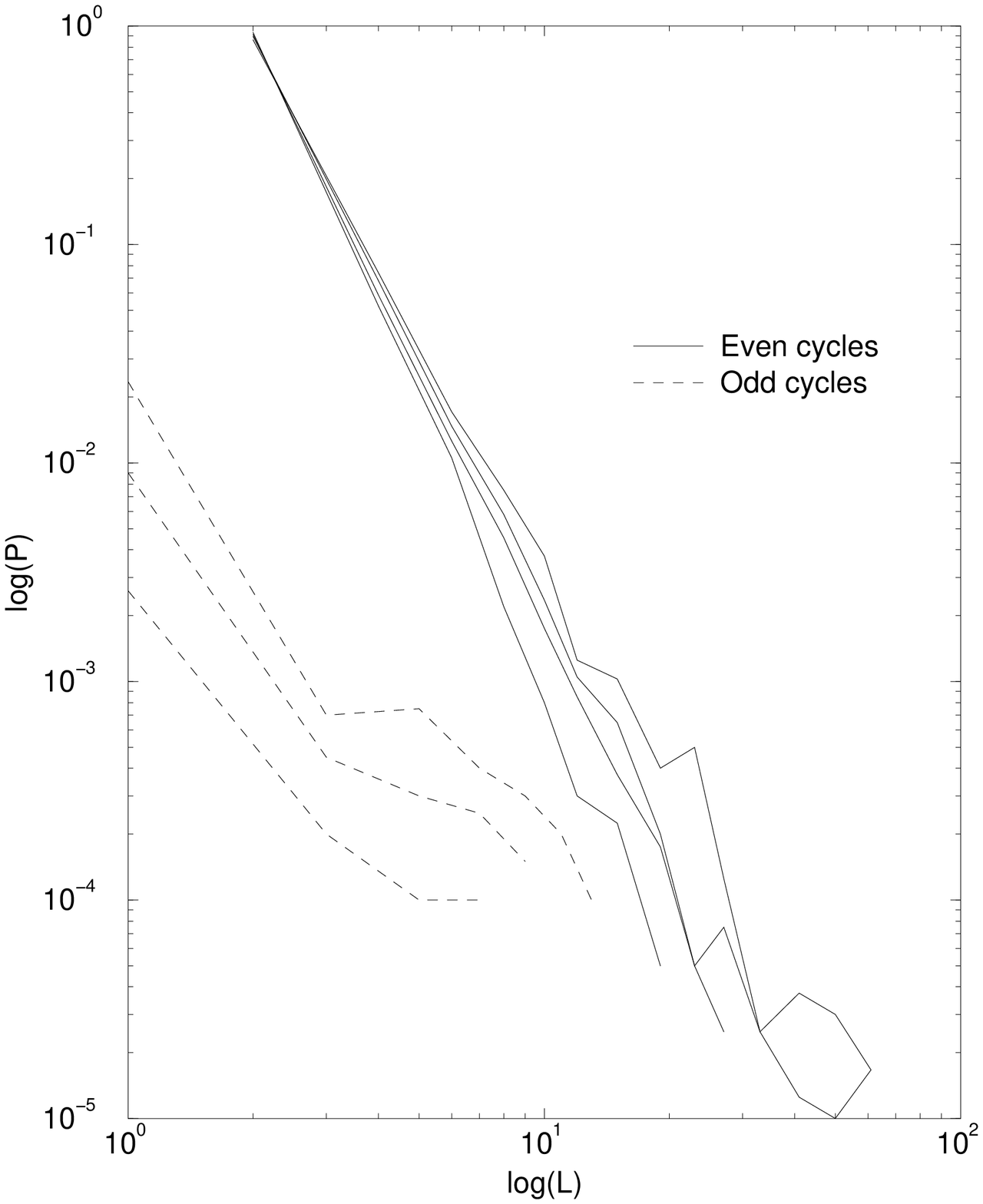,width=75truemm}
    \end{center}
  \end{minipage}

   \caption{ Probability distribution of the length of the cycles
     in the oscillatory phase: (a) $\e=0.48$, (b) $\e=0.60$. Odd and
     even lengths are shown separately. $N=32$, 48 and 64 (from top to
     bottom).}
  \label{fig_per}
\end{figure}

\begin{enumerate}

\item $\e\leq \e_c$: in this region the distribution can be divided in
  two parts: a power-law tail for short cycles, and an almost exponential
  tail for long cycles. In this part of the distribution, we found
  that the rescaled variable $L/\exp(\a_L(\e)N)$ has a well-defined
  limit distribution when $N$ increases.

As an example, we show in Fig. \ref{fig_per60} three systems of
different sizes with $\e=0.20$.

\item $\e_c<\e\leq 0.5$: in this region, the distribution is a
  power-law with exponent smaller than 2. The average value is
  dominated by the tail of the distribution. The scaling with the
  average period does not hold. As an example, we show three systems
  of different sizes with $\e=0.48$ (figure \ref{fig_per}a).

\item $\e>0.5$: in this case, the average period tends to 2. The
  distributions for the even cycles have approximately the shape of a
  power-law with an exponent larger than 2 and increasing with system size. The
  distribution of cycles of odd length is also approximately a
  power-law, with an exponent that does not decrease with system size,
  but its total weight goes to zero as $N\to\infty$. As an example, we
  show in Fig. \ref{fig_per}b four systems with $\e=0.60$.

\end{enumerate}

\subsection{Relaxation of the energy}
The function $E(t)=N^{-1}\sum_i|h_i(t)|$, where $h_i(t)$ is the local
field experienced by neuron $i$ at time $t$, is a Ljapunov function
for the symmetric system ($\e=1$). In asymmetric networks, its average
value is still a non-increasing function of time, even if it may increase
in some realizations. 

\begin{figure}
  \centerline{
    \epsfysize=8.0cm 
    \epsfxsize=10.0cm 
    \epsffile{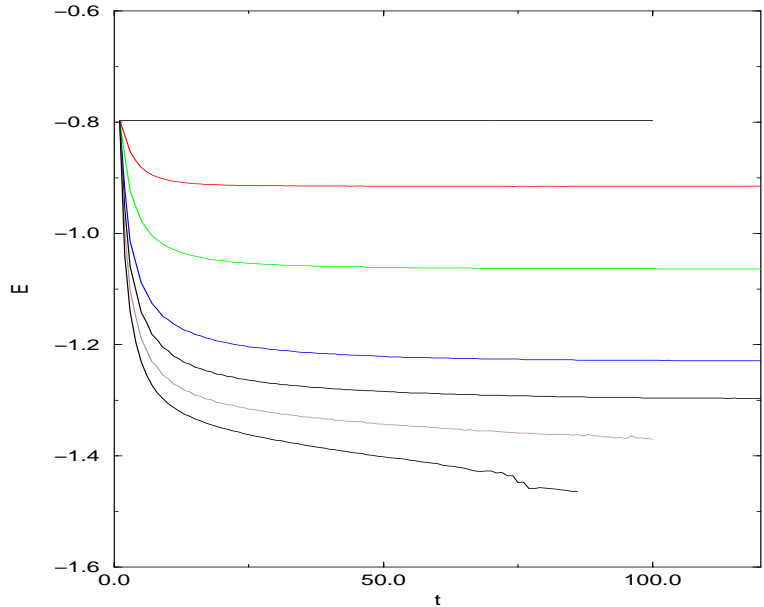}}
    \caption{$\overline{E(t)}$ on trajectories not yet closed for
    $N=256$ and $\e$ respectively equal to 0, $0.32$, $0.50$,$0.60$,
    $0.80$, $0.90$ and $1$. The energy is a decreasing function of $\e$.}
  \label{fig_ener}
\end{figure}

We show in Fig. \ref{fig_ener}a $\overline{E(t)}$ for $N=256$ and
several values of $\e$ between 0 and 1. We imposed the condition that the
trajectory is not yet closed when the energy is measured. Without
this condition, at high symmetry the trajectories ultimately
find cycles of length 2 and the energy reaches a stationary value
corresponding to the average energy of such cycles. With the opening
condition the energy decreases to lower values. The effect of the
opening condition is very small (not even significant) at small $\e$
and increases as $\e$ grows.

At $\e=0$ the energy density is constant in time, and is equal to the
expectation value of the module of a Gaussian variable,
$E(\e=0)=\sqrt{2/\pi}=0.798$. At larger $\e$, the asymptotic value of
the energy is lower and is attained later in time.
In Fig. \ref{fig_ener} we show the relaxation of the energy for
different symmetries, for systems of size $N=256$.

\begin{figure}
  \begin{minipage}{77truemm}
    \begin{center}    
      \epsfig{file=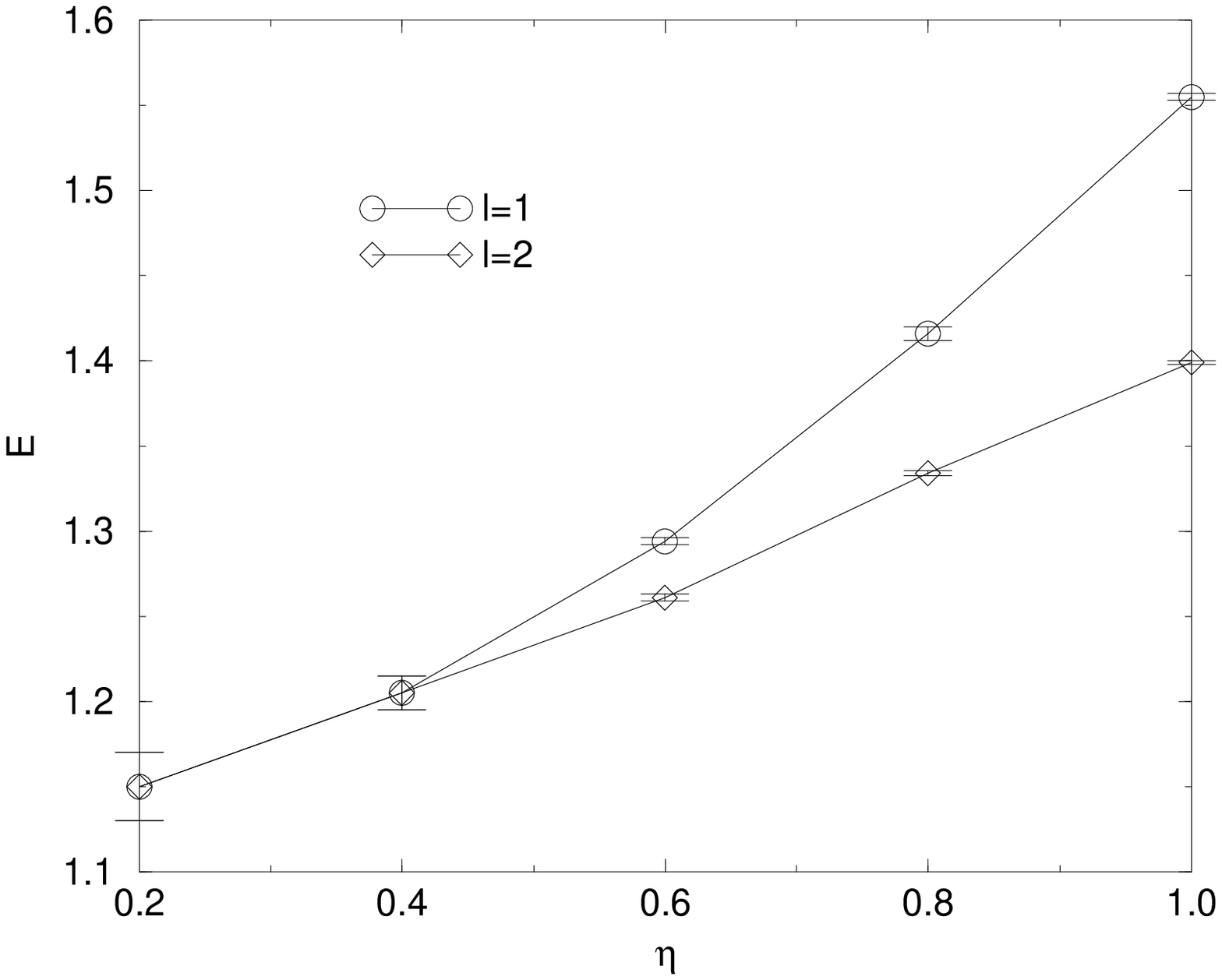,width=75truemm}
    \end{center}
  \end{minipage}
  \begin{minipage}{77truemm}
    \begin{center}    
      \epsfig{file=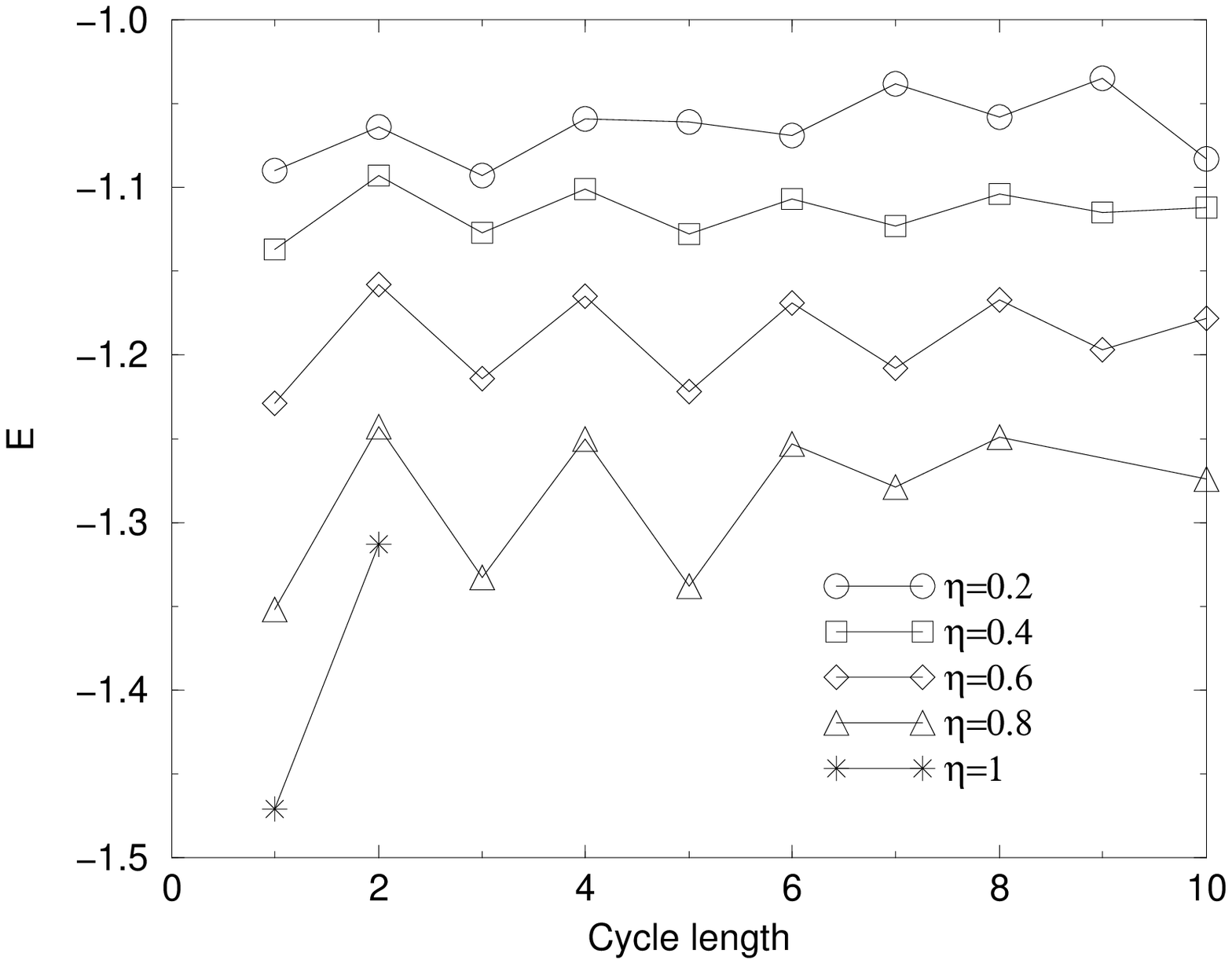,width=75truemm}
    \end{center}
  \end{minipage}

  \caption{(a) $\overline{E}$ in cycles of length 1 and 2, as a function of
      $\e$. The data are obtained extrapolating to the infinite size
      through the fit $E_N=E_\infty+AN^{-1/2}$. (b) $\overline{E}$ as
      a function of cycle length $l$ for systems with $N=64$ and
      different symmetries. For large symmetry there are no data for
      odd $l$, (except $l=1$) since such cycles are met very
      rarely. For small symmetry the small cycles are almost never
      found, and the error is very large.}
  \label{fig_en_l}
\end{figure}

At high symmetry, we observed an interesting phenomenon: the average
energy of the fixed points, that are reached with vanishing
probability as $N\to\infty$, is significantly lower than the energy
of the typical cycles of length 2. In Fig. \ref{fig_en_l}a we show,
as a function of $\e$, the infinite size extrapolation of the energy
in cycles of length 1 and of length 2 respectively (the extrapolation
was made using $N^{-1/2}$ as finite size scaling, which gives very
good fits). It can be seen
that the difference increases with $\e$. At $\e=1$ the energy of
cycles of length 2, which is also the typical energy of the parallel
dynamics, is slightly higher than the one computed in \cite{SOK} with a
Monte Carlo simulation of the infinite size dynamics. The energy of
the fixed point, on the other hand, is much lower, and its
extrapolated value, $-1.55\pm .01$, could be even lower than the
zero temperature
energy of the SK model, $E_0=-1.526$. This difference seems to reflect
a more general tendency: the cycles whose length is odd have a lower
energy than cycles of even length (figure \ref{fig_en_l}b). This
effect can be due to the fact that cycles of even length are more
rapidly found than cycles of odd length (the typical transient times
are shorter, as we expect from the closing probabilities), so the
energy has more time to decrease.

\subsection{Overlap and closing probabilities}
Though it is not the main point of this work, we present here some
numerical results about the distribution of the overlap, measured on
trajectories not yet closed (opening condition). All our data in this
section refer exclusively to the distribution of $q(t,t+l)$ subjected
to such a condition.

\begin{figure}
  \begin{minipage}{77truemm}
    \begin{center}    
      \epsfig{file=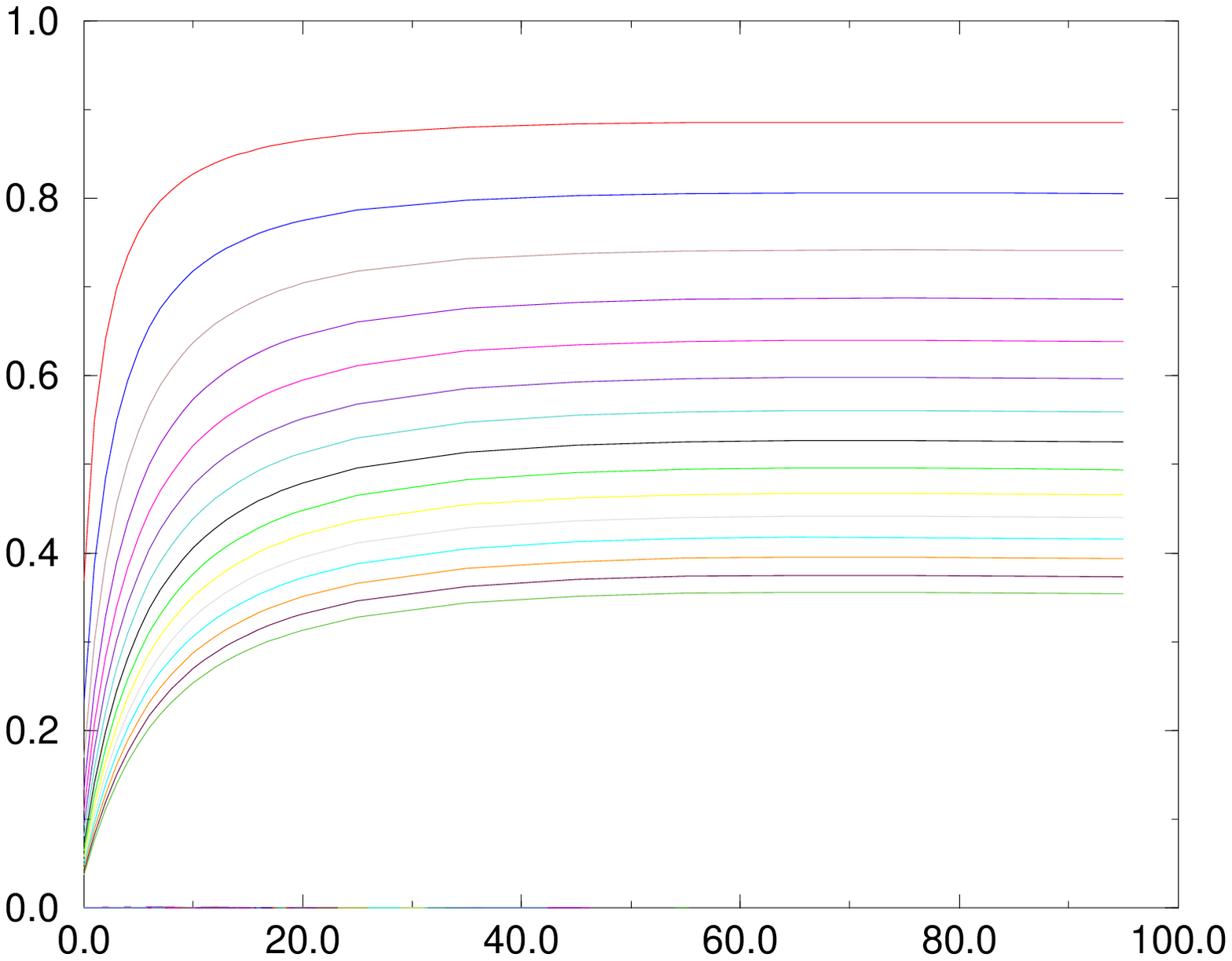,width=75truemm}
    \end{center}
  \end{minipage}
  \begin{minipage}{77truemm}
    \begin{center}    
      \epsfig{file=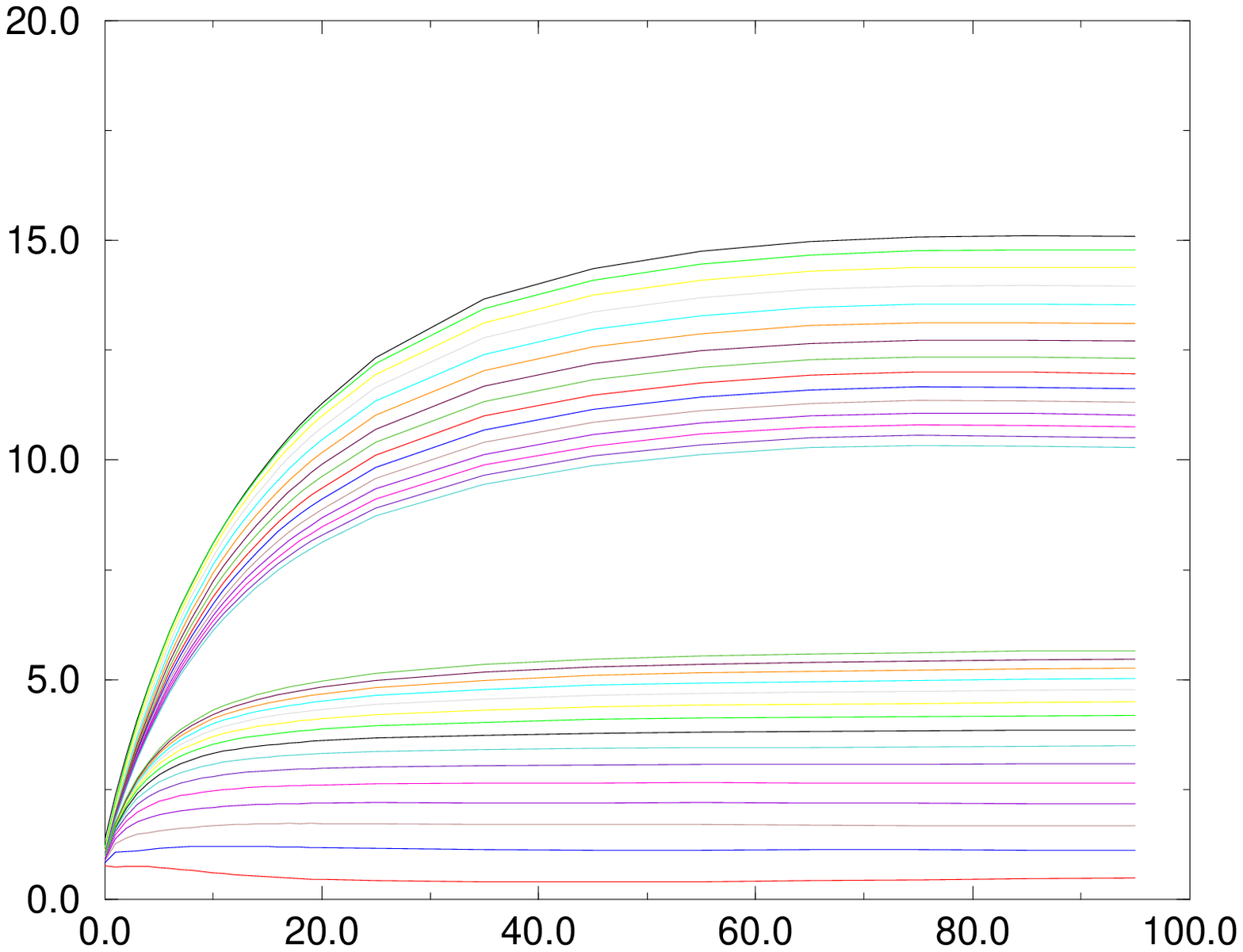,width=75truemm}
    \end{center}
  \end{minipage}

   \caption{Average value $Q(t,t+l)$ (a) and variance $NV(t,t+l)$ (b)
     of the overlap $q(t,t+l)$ over trajectories not yet closed. Here
     $\e=0.6$ and $N=128$. The different curves are for different
     values of $l$. In
     the first figure, only even values of $l$ between 2 and 30 are
     shown, from top to bottom. In the second figure the lower bundle
     of curves corresponds to even values of $l$ (and $V(t,t+l)$ is an
     increasing function of $l$), the higher bundle corresponds to odd
     $l$ (and $V(t,t+l)$ is a decreasing function of $l$).}
  \label{fig_Q}
\end{figure}

\begin{figure}
  \begin{minipage}{77truemm}
    \begin{center}    
      \epsfig{file=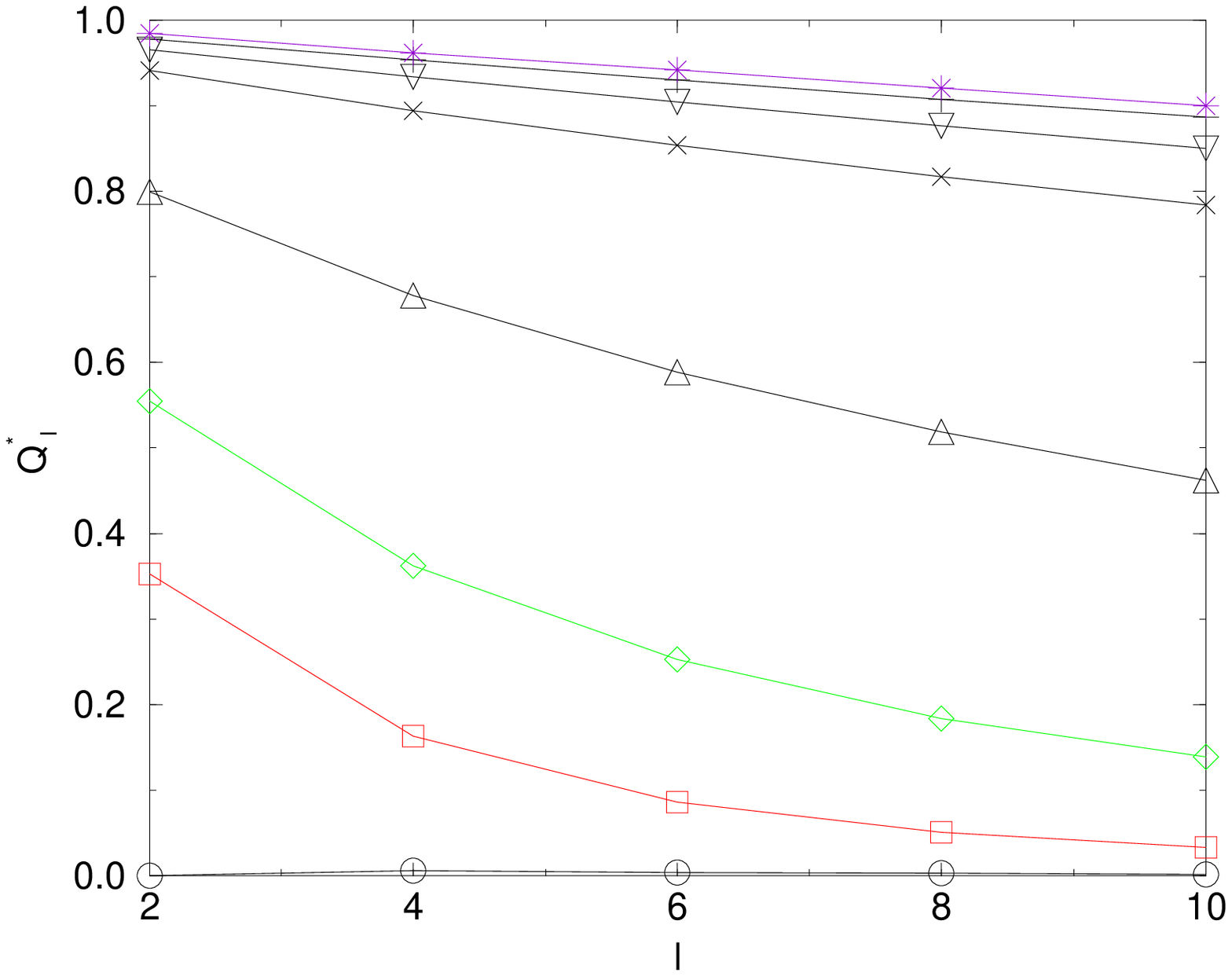,width=75truemm}
    \end{center}
  \end{minipage}
  \begin{minipage}{77truemm}
    \begin{center}    
      \epsfig{file=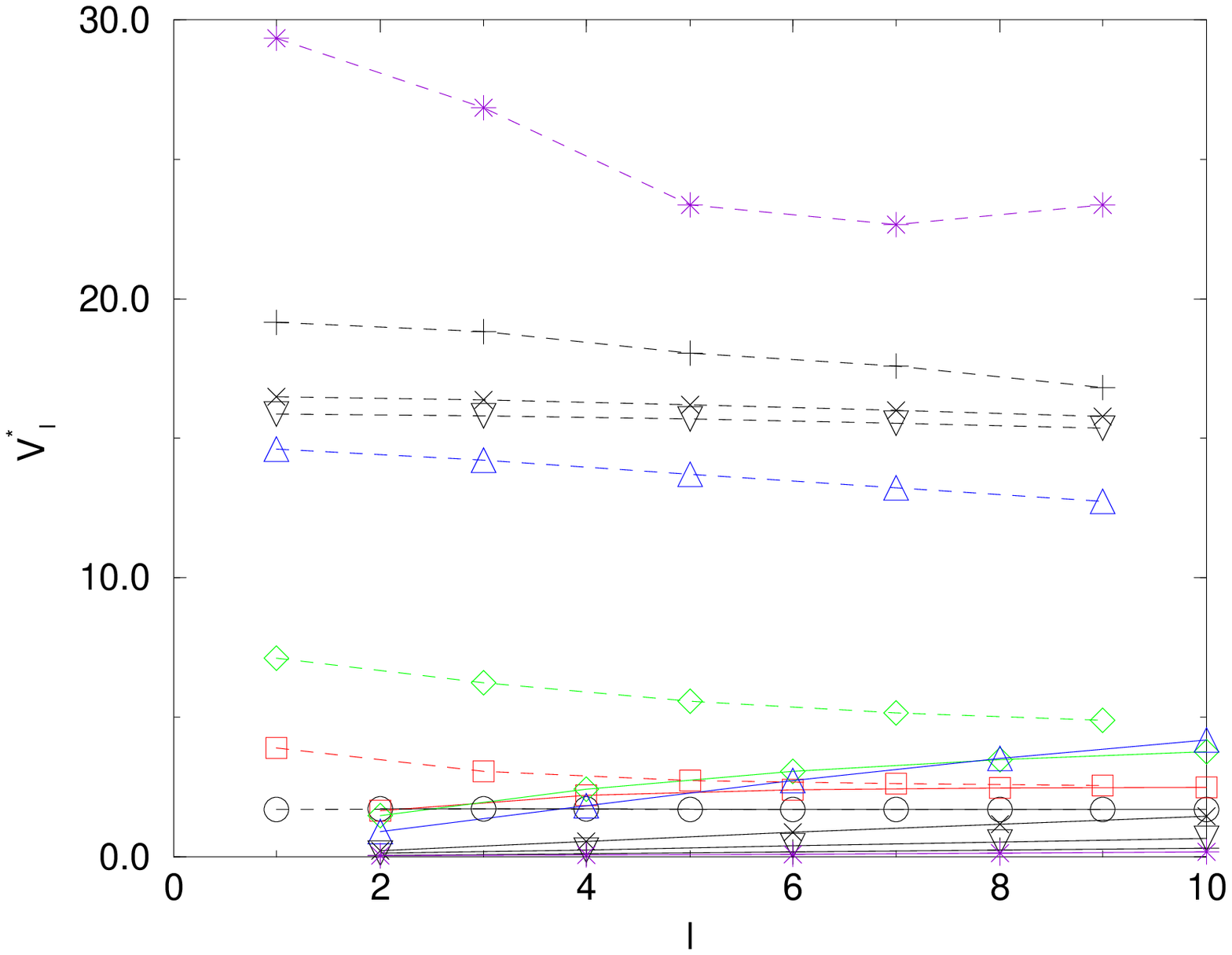,width=75truemm}
    \end{center}
  \end{minipage}

   \caption{Large time value
     of $Q(t,t+l)$ (a) and $NV(t,t+l)$ (b) as a function of $l$ for
     $N=192$ and $\e=0$, $0.2$, $0.32$, $0.5$, $0.7$,
     $0.8$, $0.9$, $1$.  In the first figure only even
     values of $l$ are shown and $\e$ grows from bottom to top. In the
     second figure even values of $l$ are represented as a dashed line
     ($\e$ grows from bottom to top) and odd values as a solid one
     ($\e$ grows from top to bottom for $l=2$). At $\e=0$ there is no
     difference between even and odd $l$.}
  \label{fig_Qe}
\end{figure}

The first figure that we show refers to the mean value and to the
variance of the distribution of $q(t,t+l)$. We show these quantities
as a function of $t$ for different values of $l$. As we noted in
section 2, the average overlap is zero due to symmetry when the time
difference $l$ is odd, so we show its value only for even $l$. For
$\e>0$, it is always a non-decreasing function of $t$, and it reaches
soon an asymptotic value (figure \ref{fig_Q}a). The fact that
$Q(t,t+l)$ is non-decreasing means that it is more and more difficult
to lose the memory of the configuration as time increases. The asymptotic
value, $Q^*_l$, is a decreasing function of $l$. We show its relaxation
for different values of the asymmetry in Fig. \ref{fig_Qe}a. It is
evident from the figure that the relaxation becomes slower and slower as $\e$
increases. Pfenning, Rieger and Schreckenberg \cite{Hei3} and
Eissfeller and Opper \cite{EO2}
observed a transition at $\e=0.825$ between a regime at high symmetry
where $Q(0,l)$ relaxes as a power-law to a non-vanishing limit value,
$Q(0,l)\approx Q_\infty+Al^{-a}$, and a regime at low symmetry where the
relaxation is exponential and the remanent magnetization vanishes,
$Q(0,l)\propto \exp(-a l)$. We expect to observe the same transition
for the asymptotic value of $Q(t,t+l)$\footnote{
We note that, due to the opening condition, this quantity has a
meaning different from the usual Edwards-Anderson order parameter,
which measures the size of an asymptotic state of the dynamics.},
but our data do not allow us to verify this point.

Figure \ref{fig_Q}b shows the variance of $q(t,t+l)$ as a
function of $t$ for different values of $l$. The variance is always an
increasing function of $t$, and reaches an asymptotic  value as $t$
increases. The cases of odd $l$ and even $l$ have to be
distinguished. The variance is larger for $l=2m+1$, and decreases as a
function of $m$. For $l=2m$ the variance is smaller, and increases as a
function of $m$. The asymptotic value of $V(t,t+l)$ is shown as a
function of $l$ (for even and odd $l$ separately) in Fig. \ref{fig_Qe}b
Since odd and even variances have an opposite behavior as a function
of $l$, the function $\tilde V(t;l)=\l(V(t,t+l)+V(t,t+l+1)\r)/2$ ($l$ odd)
shows a very small dependence on $l$. This dependence is however
systematic: $\tilde V(t;l)$ is an increasing function of $l$ when $t$
is large, and decreasing when $t$ is small. In particular, $\tilde
V(0;l)\approx 1$ (the totally random case) for every value of $\e$.
We also verified that, for $\e\simeq 1$ and $t$ large, it holds
$V(t,t+2)\propto 1-Q(t,t+2)$, relation that we used in section 2.

\begin{figure}
  \begin{minipage}{77truemm}
    \begin{center}    
      \epsfig{file=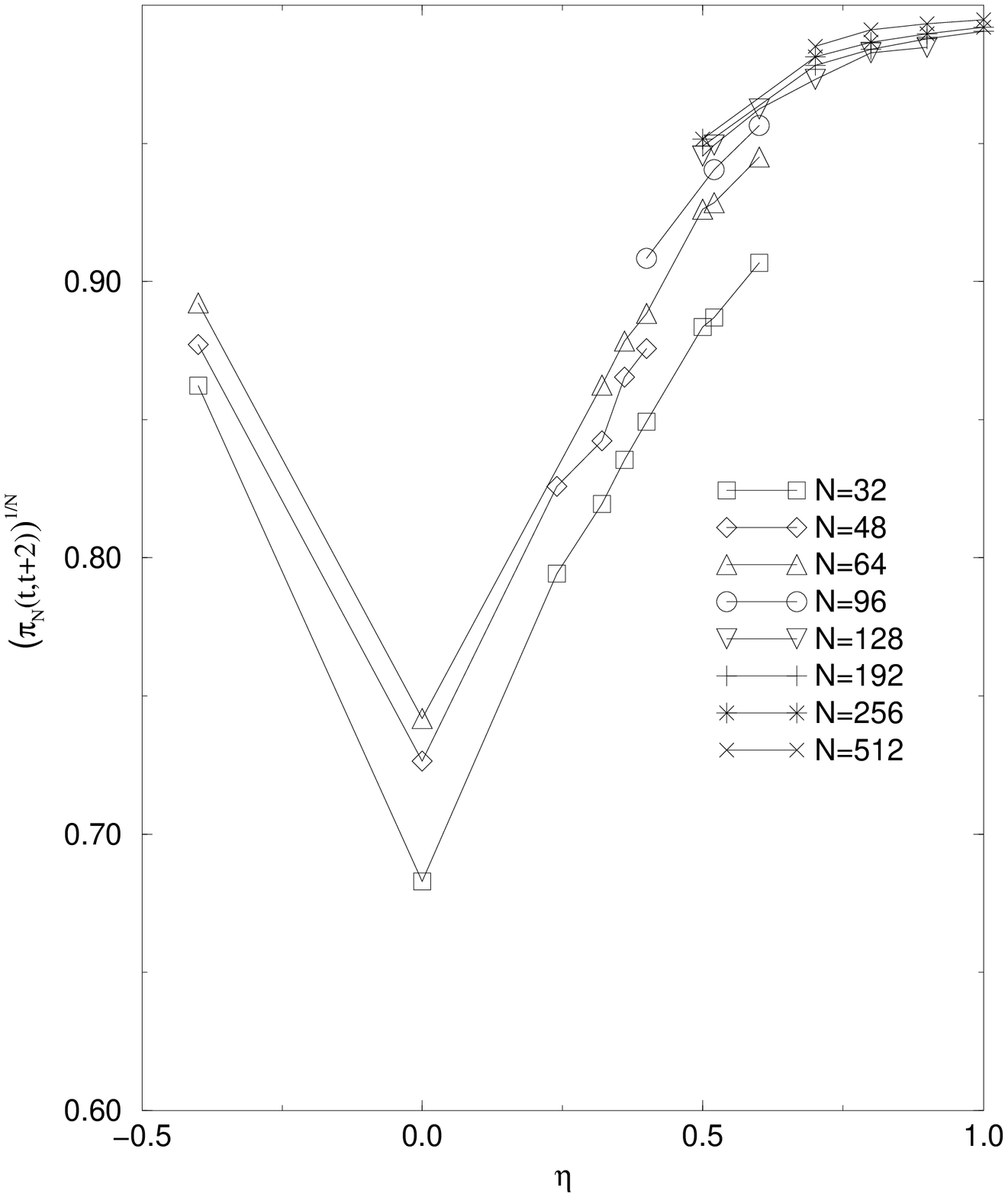,width=75truemm}
    \end{center}
  \end{minipage}
  \begin{minipage}{77truemm}
    \begin{center}    
      \epsfig{file=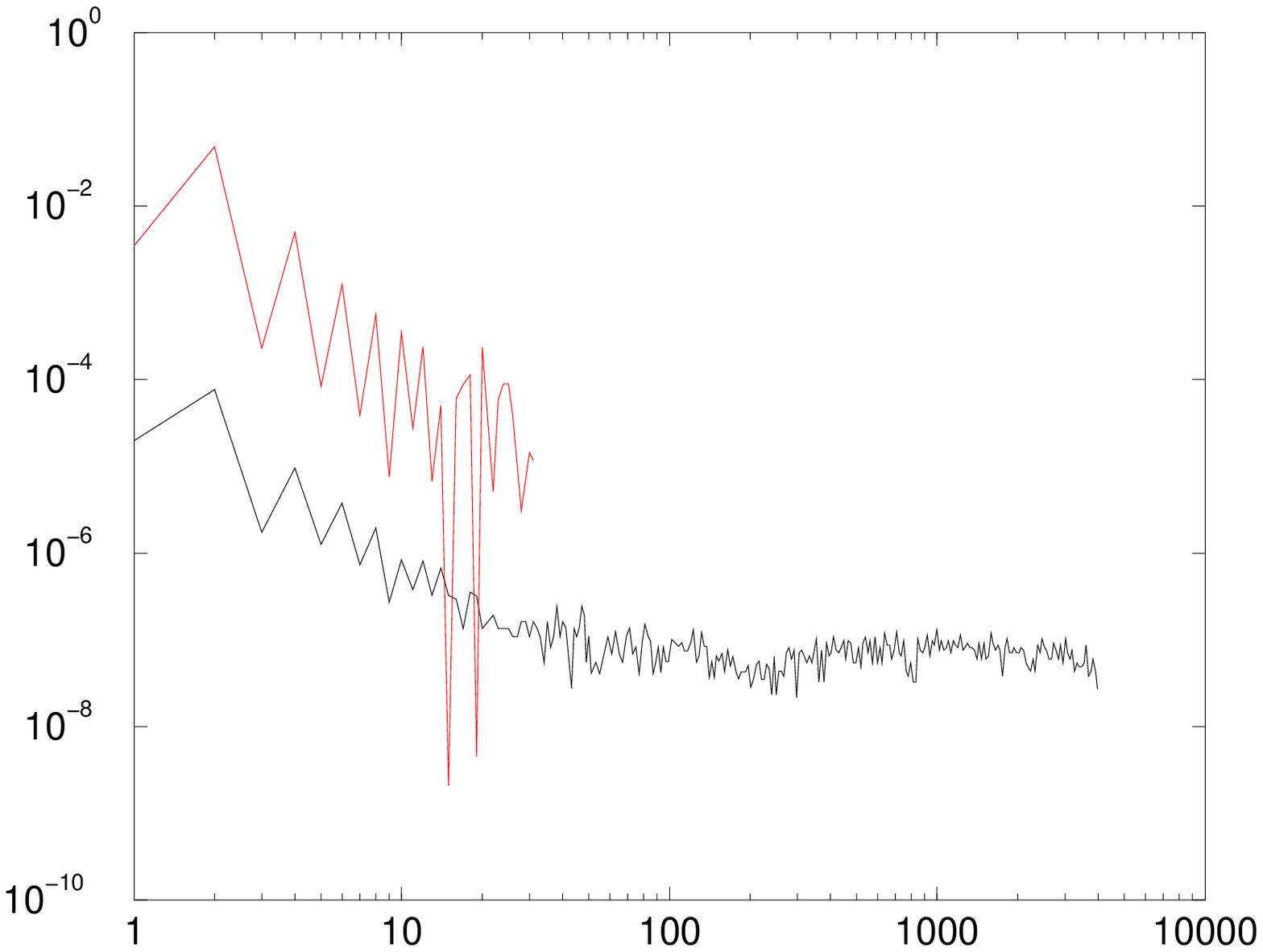,width=75truemm}
    \end{center}
  \end{minipage}

   \caption{Closing probability for cycles of length 2 as a function
  of $\e$ for systems of different sizes (a) and as a function of cycle
  length $l$, for $N=64$ and $\e=0.32$ and $0.56$ (from bottom to top).}
  \label{fig_pai}
\end{figure}

The closing probability $\pi_N(t,t+2)$ also
increases as a function of $t$ to an asymptotic value
$\pi_N^*(2;\e)\propto e^{-N\a_2(\e)}$. We plot in Fig. \ref{fig_pai}a the
behavior of $\l(\pi^*_N(2;\e)\r)^{1/N}$ as a function of $\e$ for
systems of different sizes. It appears that it converges very slowly
to a function independent of $N$, which is an even function of $\e$,
has a cusp in $\e=0$ and is concave downward. The limit value for
large $N$ is 1 for $\e=1$, because $\pi_N^*(2;\e=1)$ decreases as a
power-law of $N$, and is less than 1 for $\e<1$, indicating that the
exponential scaling is fulfilled for $\e<1$.

Before going deeper inside the analysis of the closing probabilities,
we note that all the quantities that we observed, say $Q_N(t,t+l)$,
$V_N(t,t+l)$ and $\pi_N(t,t+l)$, reach only approximately an
asymptotic value in $t$. In reality, all these quantities {\it must}
decrease with $t$, at least for time scales of the order of the
inverse of the closing probability (see section 2).
This is observed in the simulations. However, the discussion presented
above, based on stationary closing probabilities, is not modified
essentially by this fact, since the decrease of the closing
probability becomes slower and slower when system size increases: we
find, for very large $t$, and for $l=2$, $\pi_N(t,t+2)\propto
t^{-b_N}$, with $b_N\propto N^{-(1+\epsilon)}$. Thus $\pi_N(t,t+l)$
can be considered constant even on time scales exponentially
increasing with $N$, like the ones involved in the closure of the cycles.

We then report the asymptotic closing probability $\pi_N^*(l;\e)$ as a
function of $l$ for two different values of $\e$. The characteristic
oscillations for even-odd $l$ show up. Apart for that, the closing
probability is a decreasing function of $l$. It is not clear whether
it reaches a stationary value, $\exp(-N\a_\infty(\e))$, when $l$
increases. It is very difficult to measure the closing probability for
large $l$. In any case, we find that, contrarily to our hypothesis,
the value of $l$ at which $\pi_N^*(l;\e)$ seems to become stationary
increases with $N$. Thus the exponent $\a_\infty(\e)$ seems not to
exist, and in any case it can not be determined numerically.

\section{Discussion}
It is known that the deterministic dynamics of asymmetric neural
networks exhibits complex features. Among these features, we
investigated here the transition between cycles of length 2 and
``chaotic" (exponentially long) attractors, first reported in \cite{Nut}.
We found out that, though the average cycle length shows an abrupt change
at $\e_L=0.5$, the region where cycles of length 2 are stable extends
up to $\e_c=0.33\pm 0.01$. In this region nearly all the trajectories
end up in cycles of length 2. The number of these trajectories is
exponentially high and the weight of their attraction basins is
vanishingly small. At lower symmetries the probability to find a cycle
of length 2 drops abruptly to zero (though their number is still
exponentially high). The typical attractors are now cycles
exponentially long with system size, whose number increases only
proportionally to $N$. The weights of the attraction basins seem to have the
same distribution as in a Random Map model with reversal symmetry
\cite{DF,BP1}. The typical times after which attractors are reached,
on the other hand, vary exponentially with system size for every
finite asymmetry, and vary as a power-law, $Tr\propto N^{2/3}$, for
$|\e|=1$. 

All these features can be predicted with arguments
relying, through the closing probabilities, on the distribution of the
overlap in an infinite system, but the point where the transition
takes place can not be computed without a more detailed knowledge of
the overlap distribution. However, this transition does not correspond
to a transition in the infinite system. Asymmetric neural networks do
present such a transition from a phase at low symmetry where the
remanent magnetization vanishes, $Q_\infty=\lim_{l\to \infty}
Q(0,l)=0$ (loss of memory) and a phase at high symmetry where the
remanent magnetization is finite. This takes place at $\e_{RM}=0.825$
\cite{Hei3,EO2}, which is a value much larger than $\e_c=0.33$, at
which cycles of length 2 cease to be the typical attractors.

We can explain qualitatively the inclination of the trajectories to
close on cycles of length 2 as a consequence of the parallel updating
and of the symmetry of the interaction. Both these elements conspire
to create an effective ferromagnetic interaction between the spin
$\s_i(t)$ and the same spin two time steps later (for negative $\e$
the effective interaction is antiferromagnetic, and it tends to
reverse the spins after two time steps, thus resulting in cycles of
length 4). The transition takes place when the sum of the closing
probabilities of long cycles balances the one of cycles of length 2. This
balance involves both the number of cycles and the weights
of their attraction basins. Cycles of length 2 are much more than
cycles of any other length at every value of $\e>0$ (their
number grows exponentially with system size with an exponent which is
the double of the same exponent for cycles of length 1,
\cite{GRY}), but their attraction basins are vanishingly small, and,
at a certain point, very long cycles, whose number grows only linearly
with system size, represent the overwhelming majority of phase space. 

\vspace{.2cm}
The dynamics of the system is a kind of relaxation, the function $E(t)$
defined in (\ref{ener}) playing the role of an ``energy". This analogy
is exact in the symmetric system, where $E(t)$ decreases at every time step
until a cycle is reached. In the asymmetric systems $E(t)$
decreases in average, but not in every realization. The asymmetry
introduces something similar to thermal noise in the dynamics: 
the average asymptotic value of $E$ increases when decreasing $\e$.
Fixed points are states of low energy, but they are very difficult to reach
because of the competition of higher energy attractors (either cycles
of length 2 or very long cycles), which are easier to reach. 

We can ask ourselves what changes in the above description if thermal
noise is introduced. Of course thermal noise destroys the limit cycles
which are produced by the deterministic dynamics, but at low
temperature some metastable states reminiscent of the cycles of length
2 may still survive. This discussion may be put on precise basis in symmetric
networks with $\e=1$. In this case detailed balance is fulfilled with
a suitable definition of the noise, and at equilibrium the statistical
state of the system is described by a kind of Boltzmann distribution,
for which the fluctuation-dissipation theorem holds:

\be \Pr\l(\s_1\cdots\s_N\r)\propto \prod_{i=1}^N\l(e^{\b h_i}+e^{-\b
  h_i}\r), \ee
where $h_i=\sum_j J_{ij}\s_j$ is the local field experienced by spin
  $i$ \cite{Amit}. This statistical description does not hold anymore for
  asymmetric networks, for which detailed balance breaks down.

Ferraro \cite{Gio} and Scharnagl {\it et al.} \cite{SOK} computed the
asymptotic value of the energy, which in this model is defined as
$\l\la h_i \tanh(\b h_i)\r\ra$, and at $\b=\infty$ coincides with
definition \ref{ener}, for the system with $\e=1$. They used the
Monte-Carlo scheme of \cite{EO}, which gives results free of finite size
effects (even if for computational reasons there is a limit of
$t\approx 100$ on the time steps that can be reliably performed). At
temperatures larger than 0.6 the energy follows the theoretical
prediction for the SK model with remarkable accuracy, as expected in
\cite{BPR}, but at lower temperatures the energy is considerably higher
than what expected for the SK model \cite{Gio,SOK}. This may due to the
fact that the system remains trapped in metastable states
corresponding to energy minima that at $T=0$ are cycles of length
2. As we observed, the energy of these states is significantly higher
than the energy of the fixed points (which are low energy states for
the SK model) and their attraction basins cover nearly all of phase space.
We did not do direct simulations to test this interpretation, nor to
see whether these states are also found for asymmetric couplings up to
some critical value of the asymmetry, but we think that this could be
an interesting issue.

Asymmetric networks at finite temperature were recently simulated in
\cite{IM}, though with Langevin dynamics whose equilibrium
distribution, for $\e=1$, is given by the SK model. The authors found
aging effects and non-trivial overlap distributions at small but
finite asymmetry. We agree with them about the absence of aging for
$\e<0.8$: for these systems, the average overlap $Q(t,t+l)$, subject
to the opening conditions, reaches for every $l$ a
time-translation-invariant state, where it does not depend anymore on
$t$. Our data suggest that this could not hold anymore at larger
values of $\e$ (and it can be speculated that the threshold coincides
with $\e_{RM}=0.825$ at which, where the remanent magnetization is
different from zero), but the time-window that we analyzed is to small to
state anything definite about this point.

The results that we present here have also to be compared with a
recent investigation of the relaxation dynamics of the SK model at
$T=0$ \cite{Par}. Using different kinds of dynamics, defined as reluctant,
sequential and greedy, it was found, among other things, that the
typical fixed points reached have different energy densities for the
different dynamics ($E_g=-1.416$, $E_S=-1.430$ and $E_r=-1.492$).
The energy density that we found for $\e=1$, extrapolated to the
infinite size, is higher in cycles of
length 2 ($E_2=-1.399$), but it is lowest on the fixed points
($E_1=-1.55\pm .01$), and it could be even lower than the zero-temperature
energy of the SK model $E_0=-1.526$.

\vspace{0.2cm}
The transition that we investigated
presents some features similar to the one taking place
in Kauffman networks, a disordered dynamical system proposed as a
model of genetic regulation \cite{K69}. Also in that case the average
length of the attractors increases exponentially in the chaotic phase,
where the weights of the attraction basins follow the Random Map
distribution \cite{BP0}, and do not depend on system size in the frozen
phase. Nevertheless, despite the similarity between the chaotic phases
of the two models, the frozen phases are quite different. In the
frozen phase of Kauffman model the number of attractors has a finite
limit as system size increases, the average weight of the attractors
does not vanish and the transient time is also finite. Moreover, in
the infinite size limit a phase transition corresponding to the transition
for the attractors takes place, between a phase without damage
spreading and a phase where a small damage propagates to the whole
system \cite{DP}. None of these features are present in asymmetric
neural networks. While the transition in Kauffman networks is a
consequence of its finite connectivity (every element receives inputs
from exactly $K$ elements), which in turn implies that, in the frozen
phase, only a finite number of elements are relevant for the dynamics
\cite{F,BP3}, asymmetric neural networks are a system with infinite
connectivity, and their oscillatory phase shows much less order than
the frozen phase of Kauffman networks.
 
\vspace{.2cm}
Though being a system with a finite number of states, this system
shows for $\e=0$ and in the infinite size some relevant
features of chaos in continuous systems \cite{CFV}. At $\e=1$, on the
other hand, the system never loses memory of its initial
condition. At intermediate symmetries the system shows features that
are ``chaotic" and features that are ``ordered". The analogy can be
carried out also through the study of the attractors.
In discretized chaotic systems ``artificial" limit cycles are present
due to the finiteness of phase space. The attraction basins of these
cycles follow a Random Map statistics \cite{BP2}, which is in
agreement to what is observed here for $\e<0.33$. The closing time of
these cycles increases as $\eps^{-D_2/2}$ \cite{GOY,Beck,BP2} , where
$\eps$ is the discretization and $D_2$ is the correlation dimension
\cite{Gr}. This is also analogous to what is observed in
the present model for $\e<0.33$ if we identify $2^{-N}$ with $\eps^D$
and $\a_\infty/\ln 2$ with $D_2/D$ (the fact that, even in the most
``chaotic case" $\e=0$, we find $\a_\infty<\ln 2$ can be interpreted as
if in this model the correlation dimension were always less than the
dimension $D$ of the embedding space). But for $\e>0.33$ the situation
is less clear: still the closing time increases exponentially with
system size, as if there were a finite ``correlation dimension" of the
asymptotic configurations equal to $\a_2/\ln 2$, but the length of
the cycles does not increase with $N$ and the statistic of the
attraction basins is not of the Random Map type. Moreover, the number
of cycles increases exponentially with $N$. Thus neither the
analogy with a discretized chaotic system, nor the analogy with a
periodic system hold.

\section*{Acknowledgments}UB is pleased to thank Gerard Barkema, Peter
Grassberger, Normann Mousseau, Paolo Muratore Ginanneschi, Heiko
Rieger, Felix Ritort and Angelo Vulpiani for interesting discussions. 


\end{document}